\documentclass{article}

\usepackage{arxiv}

\usepackage[utf8]{inputenc}
\usepackage[T1]{fontenc}
\usepackage{url}
\usepackage{booktabs}
\usepackage{amsfonts}
\usepackage{nicefrac}
\usepackage{microtype}
\usepackage{lipsum}
\usepackage{graphicx}
\usepackage[round,authoryear]{natbib}
\usepackage{xcolor}
\usepackage{hyperref}
\usepackage{doi}
\usepackage{amsmath}
\usepackage{enumitem}

\title{How Sparse and How Noisy? Systematic Benchmarking of Inverse Physics-Informed Neural Networks for Manning Friction Estimation in Shallow Water Equations}

\author{
\href{https://orcid.org/0000-0003-0177-9733}{\includegraphics[scale=0.06]{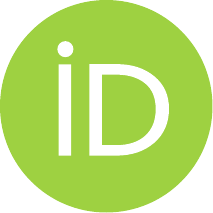}\hspace{1mm}Soheil Radfar}\thanks{Correspondence to: \texttt{sradfar@princeton.edu}}\\
Department of Civil and Environmental Engineering\\
Princeton University\\
Princeton, NJ 08544
}

\date{}

\renewcommand{\headeright}{}
\renewcommand{\undertitle}{}
\renewcommand{\shorttitle}{\textit{Radfar} 2026}

\hypersetup{
pdftitle={How Sparse and How Noisy? Systematic Benchmarking of Inverse Physics-Informed Neural Networks for Manning Friction Estimation in the Shallow Water Equations},
pdfsubject={inverse PINNs, shallow water equations, Manning friction},
pdfauthor={Soheil Radfar},
pdfkeywords={physics-informed neural networks, inverse modeling, shallow water equations, Manning friction, uncertainty},
colorlinks=true,
linkcolor=blue,
citecolor=blue,
urlcolor=blue
}

\begin{document}
\maketitle

\begin{abstract}
Physics-informed neural networks (PINNs) offer a promising framework for inverse hydrodynamic modeling because they can combine sparse observations with governing physical constraints. However, their reliability for estimating hydraulic parameters under realistic data limitations remains insufficiently characterized. This study systematically benchmarks inverse-PINN recovery of the Manning friction coefficient in the shallow water equations under controlled variations in observation sparsity, observation noise, and observed variable type. Two benchmark cases are considered: a one-dimensional MacDonald subcritical channel with an analytical steady reference solution, and a two-dimensional sloped channel with a parabolic transverse bed generated using a balanced finite-volume solver. The Manning coefficient is treated as a trainable positive scalar and recovered jointly with the flow field using a two-phase optimization strategy that first fits the available observations and then incorporates the physics residual. Results show that the two-dimensional benchmark achieves robust friction recovery, with errors below 5\% when at least 10 observations of depth and velocity are available and noise levels remain at or below 10\% of the field standard deviation. Recovery remains stable up to 20\% noise when 50 observations are used, but becomes unreliable at the sparsest setting of five observations. In contrast, the one-dimensional benchmark exhibits a persistent positive bias of approximately 15\% that is largely insensitive to observation count and noise, indicating a structural identifiability limitation rather than a data-density limitation. Observation-type ablation further shows that friction recovery degrades substantially when only depth or only velocity is observed, demonstrating that joint depth--velocity information is essential for reliable inverse identification. Overall, the results provide a reproducible benchmark for assessing when inverse PINNs can and cannot reliably estimate Manning friction from sparse and noisy shallow-water observations.
\end{abstract}

\keywords{Physics-Informed Neural Networks \and Inverse Modeling \and Shallow Water Equations \and Manning Friction \and Uncertainty}

\section{Introduction}

Numerical models based on the shallow water equations (SWEs) are central tools for simulating flood waves, river hydraulics, dam-break flows, storm surge, tsunami propagation, and coastal inundation, but their predictive skill depends strongly on uncertain parameters, boundary conditions, topography, and sparse observations \citep{Toro2001,LeVeque2002,Chaudhry2008,Bates2010,Neal2012}.
The need for reliable SWE-based modeling is particularly acute in coastal regions, where tides, storm surge, rainfall, river discharge, waves, sea-level rise, and infrastructure exposure can interact to produce nonstationary and compound flood hazards \citep{Wahl2015,Moftakhari2017,Zscheischler2018,Radfar2023WACE,Mahmoudi2025EarthsFuture,Radfar2026ERL}. Traditional calibration of hydrodynamic models is often computationally expensive because it requires repeated forward simulations, adjoint derivations, ensemble-based sampling, or manual tuning of uncertain parameters \citep{Tarantola2005,Chen2023CoastalEngineering,Someya2025GJI}.
These computational burdens become even more severe for compound flood applications because hydrologic, hydraulic, and coastal boundary processes may need to be coupled or hybridized across multiple drivers and spatial scales \citep{radfar2024nature, Radfar2026HESS}. Purely data-driven neural networks can learn complex hydrodynamic mappings, but they usually require large labeled datasets and may fail to generalize outside the training distribution when physical constraints are weak or absent \citep{Karniadakis2021,Cuomo2022,Hu2025JOE}.
Recent physics-informed and hybrid-emulation studies have shown that enforcing mass and momentum constraints can improve the physical consistency of compound-flood predictions relative to unconstrained neural networks, especially during peak flood conditions when multiple drivers interact nonlinearly \citep{Radfar2025ALPINE,Radfar2026HESS}. Physics-informed neural networks (PINNs) provide an alternative by embedding governing equations, initial conditions, boundary conditions, and sparse measurements directly into the loss function, thereby allowing the flow variables and unknown physical quantities to be inferred within a single optimization framework \citep{Raissi2019,Karniadakis2021,Lu2021,Cai2021,Cuomo2022}.
This formulation is particularly attractive for inverse hydrodynamic problems because automatic differentiation can evaluate Partial Differential Equation (PDE) residuals without constructing a problem-specific adjoint solver, while observational data can regularize the solution in regimes where the physics-only inverse problem is underdetermined \citep{Baydin2018,Jin2021NSFnet,Jagtap2020ConservativePINN,Someya2025GJI,SolverLoop2026}.
At the same time, PINNs are not automatically reliable for all flow problems because training can suffer from stiffness, spectral bias, loss imbalance, local minima, and sensitivity to the placement, density, and noise level of observations \citep{Wang2021GradientPathologies,Wang2021EigenvectorBias,Jagtap2020AdaptiveActivation,Krishnapriyan2021}.

Inverse PINNs have recently been applied to several coastal, ocean, and free-surface-flow problems where only partial observations are available.
Nearshore bathymetry and wave fields have been jointly inferred from sparse wave-height, wave-number, and depth measurements using wave-energy and dispersion constraints \citep{Holman2013,Wilson2014,Salim2021,Chen2023CoastalEngineering}.
Water-wave data assimilation and parameter identification have been demonstrated using the nonlinear Schrödinger equation, potential-flow theory, Serre--Green--Naghdi equations, and related wave models under sparse surface-elevation measurements \citep{Jagtap2022OceanEngineering,Ehlers2024Fluids,Ehlers2025PRF}.
Coastal and marine parameter estimation has also been extended to vegetation drag coefficients in Boussinesq-type wave models, seabed poroelastic parameters under wave-induced loading, offshore tsunami wavefield reconstruction from ocean-bottom pressure gauges, and submarine groundwater discharge estimation from temperature time series under tidal forcing \citep{Huang2025EAAI,Chen2026JMSE,Someya2025GJI,Frei2026LOM}.
Ocean-scale applications have further shown that PINN-based shallow-water formulations can estimate open-boundary wave amplitudes and bottom-friction coefficients from reanalysis-derived flow fields \citep{Hu2025JOE}. In parallel, coastal-flooding studies have increasingly emphasized that sparse gauges, uneven observational coverage, and spatially variable impacts remain major barriers to reliable high-tide-flooding assessment, coastal infrastructure planning, and operational road-closure decisions \citep{Mahmoudi2025EarthsFuture,Maghsoodifar2025IJDRR}. These studies demonstrate that inverse PINNs are increasingly being used not only as forward solvers, but also as physics-constrained data-assimilation and parameter-identification tools for coastal engineering and hydrology \citep{Mohammadi2025NonlinearDynamics,Someya2025GJI,Hu2025JOE}. However, most existing inverse PINN applications emphasize successful reconstruction or parameter recovery for selected cases, whereas fewer studies directly quantify how sparse and noisy observations affect the identifiability of a target hydraulic parameter \citep{Chen2023CoastalEngineering,Huang2025EAAI,Ehlers2025PRF,Frei2026LOM,Hu2025JOE}.

Among the uncertain parameters in shallow-water modeling, bottom friction is one of the most influential because it controls momentum dissipation, wave attenuation, flood-wave propagation, flow velocity, and water-surface gradients \citep{Chow1959,Arcement1989,Cunge1980,Chaudhry2008}.
In many practical models, friction is represented through a Manning coefficient that is assigned from land-cover classes, channel material, empirical tables, calibration against water levels, or expert judgment \citep{Arcement1989,Horritt2002,Neal2012,Medeiros2012,Sampson2015}.
Even when high-quality land-cover or roughness datasets are available, the effective Manning coefficient used in a depth-averaged hydrodynamic model may not correspond directly to a local physical surface property because it absorbs unresolved vegetation, bedforms, subgrid topography, wetting--drying effects, numerical diffusion, and scale-dependent hydraulic resistance \citep{Medeiros2012,Neal2012,Dyakonova2018,Sampson2015, wang2026unraveling}. This scale dependence is especially important in coastal floodplains, roads, wetlands, and urban transition zones, where small differences in depth and velocity can control inundation impacts, transportation disruption, and vehicle-stability thresholds \citep{Bates2010,Neal2012,Maghsoodifar2025IJDRR}.
This makes friction estimation a natural inverse problem in the SWEs, especially when the available information consists of sparse water-depth or velocity observations rather than spatially complete flow fields \citep{Tarantola2005,Raissi2019,Hu2025JOE}. The problem is also connected to broader coastal risk practice because nonstationary extremes, dependence among marine variables, high-tide flooding, and evolving tidal dynamics all influence the hydraulic conditions under which friction-sensitive flood models are used for design and risk assessment \citep{Radfar2023WACE,Radfar2023OceanEngineering,Mahmoudi2025EarthsFuture,Radfar2026ERL}.

Despite the growing number of inverse PINN studies in coastal and free-surface hydraulics, the reliability of Manning-friction recovery under controlled observation limitations remains insufficiently characterized.
Existing studies often demonstrate successful recovery of a specific parameter or flow field for a selected case, but they rarely provide a systematic map of how estimation error changes as a function of observation sparsity, observation noise, observed variable type, and spatial dimension \citep{Chen2023CoastalEngineering,Huang2025EAAI,Ehlers2025PRF,Frei2026LOM,Hu2025JOE}.
This limitation is important because a successful inverse PINN result under dense, clean, synthetic observations does not necessarily imply that the same method will remain identifiable when only a few noisy gauges or velocity samples are available \citep{Wang2021GradientPathologies,Krishnapriyan2021,Chen2026JMSE}.
For Manning-friction estimation, this issue is especially critical because different combinations of friction, boundary forcing, bathymetry, and initial conditions can produce similar water-level responses, leading to equifinality and non-unique inverse solutions \citep{Beven2006,Tarantola2005,Horritt2002}. Such identifiability concerns are not merely numerical: coastal flood studies increasingly show that uncertainty in drivers, boundary conditions, tidal behavior, sea-level trends, and compound interactions can propagate into risk metrics and infrastructure decisions \citep{wang2026unraveling,Radfar2023OceanEngineering,Mahmoudi2025EarthsFuture,Radfar2026HESS,Radfar2026ERL}.
A reproducible benchmark that explicitly quantifies the trade-off between sensor density, noise magnitude, and friction-recovery accuracy is therefore needed before inverse PINNs can be interpreted as robust calibration tools for practical SWE applications \citep{Raissi2019,Karniadakis2021,Cai2021,Cuomo2022}. Such a benchmark also provides a necessary bridge between physics-informed learning and applied coastal decision-making, because practical applications rarely have dense, noise-free observations across the full floodplain \citep{Mahmoudi2025EarthsFuture,Maghsoodifar2025IJDRR,Radfar2025ALPINE}.

This study addresses this gap through a systematic benchmarking framework for inverse-PINN recovery of the Manning friction coefficient in one- and two-dimensional shallow-water flows.
The benchmark is designed around canonical SWE test cases with known reference solutions, allowing the recovered friction coefficient to be evaluated against a controlled ground truth rather than against an uncertain field-calibrated value.
The experiments quantify how friction-recovery accuracy varies across a structured grid of observation sparsity and observation noise, and they compare the information content of different observation types, including water depth, velocity, and combined depth--velocity measurements.
This design directly targets the practical question of how much and what type of information is required before inverse PINN-based friction estimates become trustworthy for SWE applications \citep{Raissi2019,Karniadakis2021,Hu2025JOE}. The study reports multi-seed statistics, error surfaces, and practical identifiability thresholds to distinguish robust recovery regimes from configurations where inverse PINN estimates become unstable or non-unique. The resulting benchmark does not claim to provide a universal calibration law for all hydraulic systems; instead, it provides a reproducible and interpretable reference for evaluating when inverse PINNs can and cannot recover Manning friction in the SWEs under controlled data limitations. By explicitly connecting inverse PINN identifiability to observation sparsity and noise, the study complements emerging coastal engineering efforts that combine physics-based modeling, machine learning, and sparse observations to improve flood prediction, attribution, infrastructure planning, and risk assessment \citep{Radfar2025ALPINE,Radfar2026HESS,Maghsoodifar2025IJDRR,Mahmoudi2025EarthsFuture}.

\section{Materials and Methods}
\label{sec:methods}

This section describes the inverse PINN framework used to recover the Manning roughness coefficient from sparse and noisy SWE observations. We state the governing equations and the two benchmark cases (Sections~\ref{subsec:swe}--\ref{subsec:benchmarks}), the inverse PINN formulation (Section~\ref{subsec:pinn}), the two-phase training procedure (Section~\ref{subsec:training}), the design choices that were established by diagnostic experiments (Section~\ref{subsec:design}), and the benchmarking protocol (Section~\ref{subsec:benchmark}).

\subsection{Governing equations}
\label{subsec:swe}

In one spatial dimension, the conservative shallow water equations with Manning bottom friction read
\begin{align}
\frac{\partial h}{\partial t} + \frac{\partial (h u)}{\partial x} &= 0, \label{eq:swe1d_mass} \\
\frac{\partial (h u)}{\partial t} + \frac{\partial}{\partial x}\!\left( h u^{2} + \tfrac{1}{2} g h^{2} \right) &= -g h \frac{\partial z_{b}}{\partial x} - g h\, S_{f}, \label{eq:swe1d_mom}
\end{align}
where $h(x,t)$ is the water depth, $u(x,t)$ is the depth-averaged velocity, $z_{b}(x)$ is the bed elevation, $g$ is gravitational acceleration, and the Manning friction slope is
\begin{equation}
S_{f} = \frac{n^{2}\, u\, |u|}{h^{4/3}}.
\label{eq:manning_1d}
\end{equation}
The Manning coefficient $n$ is the parameter to be recovered. In two spatial dimensions, the SWEs for $h(x,y,t)$, $u(x,y,t)$, $v(x,y,t)$ are
\begin{align}
\frac{\partial h}{\partial t} + \frac{\partial (h u)}{\partial x} + \frac{\partial (h v)}{\partial y} &= 0, \label{eq:swe2d_mass} \\
\frac{\partial (h u)}{\partial t} + \frac{\partial}{\partial x}\!\left( h u^{2} + \tfrac{1}{2} g h^{2} \right) + \frac{\partial (h u v)}{\partial y} &= -g h \frac{\partial z_{b}}{\partial x} - g h\, S_{fx}, \label{eq:swe2d_momx}\\
\frac{\partial (h v)}{\partial t} + \frac{\partial (h u v)}{\partial x} + \frac{\partial}{\partial y}\!\left( h v^{2} + \tfrac{1}{2} g h^{2} \right) &= -g h \frac{\partial z_{b}}{\partial y} - g h\, S_{fy}, \label{eq:swe2d_momy}
\end{align}
with
\begin{equation}
S_{fx} = \frac{n^{2}\, u\, \sqrt{u^{2}+v^{2}}}{h^{4/3}}, \qquad
S_{fy} = \frac{n^{2}\, v\, \sqrt{u^{2}+v^{2}}}{h^{4/3}}.
\label{eq:manning_2d}
\end{equation}
Both benchmarks below are constructed at steady state, so the time derivatives on the left-hand sides vanish in the reference solutions; the PINN residuals retain the full form for generality.

\subsection{Benchmark problems}
\label{subsec:benchmarks}

To evaluate parameter recovery against a controlled ground truth, we use two benchmarks: a one-dimensional analytical case and a two-dimensional case in which a steady reference field is produced by a well-balanced finite-volume solver. In both, the Manning coefficient that enters the data generation is treated as the target value $n_{\mathrm{true}}$ against which the recovered $\hat{n}$ is compared.

\paragraph{1D MacDonald subcritical channel.}
The first benchmark follows the construction of MacDonald et al.~\cite{MacDonald1997}, which is also catalogued in the SWASHES library~\cite{Delestre2013}. We prescribe a smooth subcritical depth profile,
\begin{equation}
h(x) = 0.5 + 0.1\, \sin\!\left(\frac{\pi x}{L}\right), \qquad x \in [0, L],
\label{eq:hprofile}
\end{equation}
with $L = 1000$\,m, constant unit discharge $q = 0.5$\,m\textsuperscript{2}/s, and target $n_{\mathrm{true}} = 0.02$. The bed elevation $z_{b}(x)$ is then \emph{back-solved} so that the steady momentum equation is exactly satisfied for the chosen $(h, q, n_{\mathrm{true}})$:
\begin{equation}
\frac{d z_{b}}{d x}
= -\frac{1}{g h}\frac{d}{dx}\!\left( \frac{q^{2}}{h} + \tfrac{1}{2} g h^{2} \right)
- S_{f},
\qquad
z_{b}(x) = \int_{0}^{x} \frac{d z_{b}}{d x'}\, d x'.
\label{eq:back_solve_z}
\end{equation}
This construction is necessary because a flat-bed assumption with the depth profile in Eq.~\eqref{eq:hprofile} does \emph{not} correspond to any constant-$n$ steady solution; back-solving the bed makes the resulting $(h, u, z_{b})$ a self-consistent steady solution of Eqs.~\eqref{eq:swe1d_mass}--\eqref{eq:swe1d_mom} for the prescribed $n_{\mathrm{true}}$. The flow remains subcritical throughout, with Froude number $\mathrm{Fr} = u/\sqrt{g h} < 1$. The dataset is sampled on a uniform grid of $N = 501$ points.

\paragraph{2D sloped channel with parabolic transverse bed.}
The second benchmark is a steady two-dimensional open-channel flow over a bed of the form
\begin{equation}
z_{b}(x, y) = -S_{0}\, x + h_{y}\!\left( \frac{2 y}{W} \right)^{2},
\qquad
(x, y) \in [0, L_{x}] \times [-W/2,\, W/2],
\label{eq:bed_2d}
\end{equation}
which combines a gentle streamwise slope $S_{0}$ with a parabolic transverse cross-section of amplitude $h_{y}$. We use $L_{x} = 2000$\,m, $W = 400$\,m, $S_{0} = 2 \times 10^{-3}$, $h_{y} = 0.3$\,m, unit inflow discharge $q_{\mathrm{in}} = 1.0$\,m\textsuperscript{2}/s, and $n_{\mathrm{true}} = 0.02$. The streamwise slope drives a friction-balanced normal-depth flow, while the transverse curvature produces depth-varying friction and a non-trivial cross-channel velocity component, so that all three state variables $h$, $u$, $v$ carry information about $n$.

This case does not admit a closed-form solution. The reference state is therefore generated by a dedicated finite-volume SWE solver implemented for this study, using an HLL Riemann solver at cell interfaces, the Audusse hydrostatic reconstruction~\cite{audusse2004fast} for a well-balanced treatment of the bed-slope source term, a point-implicit update for the Manning friction, and forward-Euler time stepping under a Courant–Friedrichs–Lewy (CFL) constraint. The solver is integrated from a uniform normal-depth initial condition with Dirichlet inflow on the west boundary, zero-gradient outflow on the east boundary, and free-slip walls on the north and south boundaries, until the residual of the conservative state drops below $10^{-8}$. The benchmark uses a grid of $121 \times 41$ cells; the wet interior used for training and evaluation excludes one cell along each boundary to remove ghost-cell artefacts. We verified self-consistency of the resulting dataset by back-solving $n^{2}$ pointwise from the discrete momentum balance in the central interior; the median back-solved value matches $n_{\mathrm{true}}^{2}$ within a few percent, confirming that the steady reference is identifiable in $n$.

\subsection{Inverse PINN formulation}
\label{subsec:pinn}

The PINN approximates the SWE state with a fully connected feed-forward neural network. In one dimension, the network maps $(x,t) \mapsto (\hat{h}, \hat{u})$; in two dimensions, since the reference is steady, it maps $(x,y) \mapsto (\hat{h}, \hat{u}, \hat{v})$. We use $\tanh$ activations and a width and depth of eight hidden layers of twenty neurons each, matching the forward-PINN baseline from which this codebase was extended. Inputs are min-max scaled to $[0,1]$ on each axis, and outputs are scaled by their per-component maximum absolute value; gradients in the PDE residuals are corrected for these scalings.

The Manning coefficient is treated as a single unknown trainable scalar. To enforce $n > 0$ throughout training without hard-clipping the gradient, we parameterize
\begin{equation}
n = \mathrm{softplus}(n_{\mathrm{raw}}),
\label{eq:n_param}
\end{equation}
where $n_{\mathrm{raw}}$ is an unconstrained learnable variable initialized so that $\mathrm{softplus}(n_{\mathrm{raw}}) = n_{\mathrm{init}}$. The variable $n_{\mathrm{raw}}$ is explicitly appended to the model's list of trainable variables; in our Keras implementation this requires overriding the \texttt{trainable\_variables} property, because scalar parameters are not automatically registered with the optimizer.

The total loss is a weighted sum of an observation term and a PDE-residual term,
\begin{equation}
\mathcal{L}(\boldsymbol{\theta}, n_{\mathrm{raw}})
= \lambda_{\mathrm{obs}}\, \mathcal{L}_{\mathrm{obs}}
+ \lambda_{\mathrm{eq}}\, \mathcal{L}_{\mathrm{eq}},
\label{eq:total_loss}
\end{equation}
where $\boldsymbol{\theta}$ are the network weights and biases. The observation loss is the mean squared error between the network prediction and the (possibly noisy) measurements at a sparse set of observation locations:
\begin{equation}
\mathcal{L}_{\mathrm{obs}} = \frac{1}{N_{\mathrm{obs}}}
\sum_{i=1}^{N_{\mathrm{obs}}}
\sum_{q \in \mathcal{Q}} \big( \hat{q}(\boldsymbol{x}_{i}) - q^{\mathrm{obs}}_{i} \big)^{2},
\label{eq:obs_loss}
\end{equation}
where $\mathcal{Q} \subseteq \{h, u\}$ in 1D and $\mathcal{Q} \subseteq \{h, u, v\}$ in 2D, and the default training configuration uses the full set. The residual loss is the mean squared residual of Eqs.~\eqref{eq:swe1d_mass}--\eqref{eq:swe1d_mom} (in 1D) or Eqs.~\eqref{eq:swe2d_mass}--\eqref{eq:swe2d_momy} (in 2D), evaluated at $N_{c}$ collocation points sampled from the wet domain. Flux derivatives are taken by automatic differentiation through the network; the bed-slope source term $g h \nabla z_{b}$ is evaluated using bed-slope components pre-computed from the benchmark bathymetry by central differences and supplied alongside the collocation coordinates. We pre-compute $\nabla z_{b}$ rather than differentiate it through the network because $z_{b}$ is a known input to the inverse problem, not an output to be inferred.

\subsection{Two-phase training procedure}
\label{subsec:training}

The joint variables $(\boldsymbol{\theta}, n_{\mathrm{raw}})$ are optimized with the Adam optimizer in two sequential phases. In \textit{Phase A}, the physics weight is set to zero ($\lambda_{\mathrm{eq}} = 0$) and the network is trained exclusively against the observation loss. In \textit{Phase B}, $\lambda_{\mathrm{eq}}$ is restored to its target value and the network weights and $n_{\mathrm{raw}}$ are jointly optimized against the full loss in Eq.~\eqref{eq:total_loss}. Phase A and Phase B each receive half of the total Adam budget. No second-order refinement is applied after Adam; the rationale is discussed in Section~\ref{subsec:design}.

The two-phase schedule is physically motivated. When the physics residual is activated from the first training step, the inverse problem is ill-posed in the early optimization because $n_{\mathrm{raw}}$ couples to a flow field that the network has not yet learned to approximate. In this regime, the optimizer can reduce the residual by driving $n$ toward zero, which trivially eliminates the friction source term in Eqs.~\eqref{eq:manning_1d} and \eqref{eq:manning_2d}, while simultaneously distorting the predicted flow field to absorb the resulting imbalance. This failure mode produces a training run with monotonically decreasing losses that nonetheless converges to a physically spurious parameter value. Pre-conditioning the network on observations alone in Phase A resolves this degeneracy by establishing a flow field broadly consistent with the data before the physics residual becomes active, so that the subsequent joint optimization drives $n_{\mathrm{raw}}$ toward the physically correct value rather than toward a loss-reducing but physically meaningless attractor.

A subtle but consequential implementation constraint arises from the phase transition. Because the loss weights change between phases, they must remain dynamically readable by the optimizer at every training step. In automatic-differentiation frameworks, wrapping the training step in a compiled graph function causes any quantities treated as compile-time constants to be fixed at their values from the first invocation. If the loss weights are implemented in this way, the Phase B switch has no effect on the actual computation, and the network continues to optimize the Phase A objective throughout both phases. To avoid this silent failure, the loss weights must be implemented as mutable state variables that are updated between phases, so that the compiled graph reads their current values at each step rather than their values at trace time. This implementation detail leaves no visible signature in the loss curves, making it a particularly difficult failure mode to diagnose without explicit inspection of the gradient flow to $n_{\mathrm{raw}}$.

\subsection{Design choices and their empirical basis}
\label{subsec:design}

The final configuration was arrived at through a sequence of diagnostic experiments in which each hyperparameter was varied systematically, and the gradient flow to the friction parameter was monitored explicitly at each stage.

\paragraph{Network architecture.}
The architecture of eight $\tanh$-activated hidden layers of twenty neurons each was inherited from the forward PINN baseline and retained after confirming that it reproduces both benchmark reference fields to below one percent relative $L_{2}$ error in a supervised forward-only configuration. Pilot runs with wider networks (thirty neurons) and shallower networks (six layers) did not improve inverse recovery and increased training cost; the reported architecture is the smallest configuration that supports accurate forward reconstruction of both benchmarks.

\paragraph{Optimizer selection and the exclusion of second-order refinement.}
The forward-PINN baseline from which this work was extended follows the convention of Raissi et al.~\cite{Raissi2019} and uses a first-order Adam warm-up followed by an L-BFGS quasi-Newton refinement. An initial implementation of the inverse PINN adopted the same schedule. Diagnostic runs revealed that the Adam phase consistently recovered $n$ to within two to four percent of the true value, but the subsequent L-BFGS stage reliably degraded the recovered $n$ to errors exceeding seventy percent, even while reducing both the observation loss and the physics residual further. Trajectory inspection confirmed that L-BFGS was exploiting a near-null direction in the joint parameter-and-network landscape: because the network has far more degrees of freedom than the number of observations, there exists a family of solutions in which an incorrect friction coefficient is compensated by a distorted flow field such that both losses decrease simultaneously. L-BFGS, being a precise line-search method, traverses this direction efficiently; Adam, with its bounded per-step displacement and gradient noise, does not. All reported experiments therefore use Adam alone, without any second-order refinement stage.

\paragraph{Loss weights.}
An initial implementation weighted the physics residual by a factor of $10^{6}$, in conjunction with a stop-gradient-based normalization of the loss components that is common in forward-PINN practice. Diagnostic inspection of the gradient magnitude reaching the friction parameter showed that this normalization reduced the gradient to machine zero: the normalization constant was computed from a quantity that was stopped from contributing gradients, so the friction parameter received no meaningful update despite monotonically decreasing losses. Removing the normalization restored gradient flow to the friction parameter. Without normalization, the observation and physics loss components are of comparable magnitude at the operating point of the trained network, and equal weights $\lambda_{\mathrm{obs}} = \lambda_{\mathrm{eq}} = 1.0$ were found to produce stable recovery across all benchmark configurations. The earlier large weight was a calibration artefact of the now-removed normalization scheme and is not appropriate once raw unscaled losses are used.

\paragraph{Training budget.}
The total Adam budget was varied across $\{1000, 2000, 3000, 4000\}$ steps for both benchmarks. In the 1D case, extending training beyond 1000 steps produced negligible additional movement in the recovered friction coefficient relative to the residual bias from $n_{\mathrm{true}}$, and the shorter budget was therefore adopted (the residual descent visible in Figure~\ref{fig:trajectory_1d} is small in absolute terms and is interpreted in Section~\ref{subsec:discussion} as a structural identifiability bias rather than an artefact of undertraining). In the 2D case, the larger spatial domain and the presence of the transverse velocity component required a longer optimization to establish a consistent flow field in Phase~A; a budget of 3000 steps was found to be sufficient for the friction coefficient to reach its asymptotic regime, while 1000 steps produced incomplete Phase~A convergence and consequently noisier Phase~B recovery. The learning rate was fixed at $10^{-3}$ throughout, which is the Adam default and was found to be adequate without decay scheduling for the run lengths used.

\paragraph{Phase split.}
Within the two-phase schedule, equal allocation of the total budget between Phase A and Phase B was adopted after testing asymmetric splits. Allocating fewer than half the steps to Phase A left the flow field insufficiently pre-conditioned and caused Phase B to recover a biased $n$; extending Phase A beyond half the budget did not improve the Phase B outcome but reduced the number of steps available for joint optimization.

\subsection{Benchmark design}
\label{subsec:benchmark}

The benchmark is organized as a structured sweep over three axes that together control the difficulty of the inverse problem:
\begin{enumerate}
    \item \textbf{Observation sparsity.} The number of observation points $N_{\mathrm{obs}}$ is varied over $\{5, 10, 20, 50\}$, spanning from severely undersampled to moderately dense relative to the reference grid.
    \item \textbf{Observation noise.} Independent Gaussian noise with standard deviation $\sigma$ equal to a fixed fraction of the standard deviation of each observed field is added to each measurement. The relative noise level is varied over $\{0\%, 5\%, 10\%, 20\%\}$.
    \item \textbf{Observation type (1D only).} The observed variable set $\mathcal{Q}$ in Eq.~\eqref{eq:obs_loss} is varied across $\{h\}$, $\{u\}$, and $\{h, u\}$ at a fixed point on the grid ($N_{\mathrm{obs}} = 20$, $\sigma = 0$) to isolate the information content of each observation type for friction recovery.
\end{enumerate}

For each configuration in the sparsity--noise grid, training is repeated over $n_{\mathrm{seeds}} = 5$ independent random seeds controlling network initialization, observation-point sampling, and noise realizations. Recovery is summarized by the relative error
\begin{equation}
\varepsilon_{n} = \frac{|\hat{n} - n_{\mathrm{true}}|}{n_{\mathrm{true}}} \times 100\%,
\label{eq:n_error}
\end{equation}
reported as a seed-mean and seed-standard-deviation across runs. Multi-seed statistics distinguish robust recovery regimes from configurations in which the inverse problem becomes practically non-identifiable, in the sense that seed-to-seed variability of $\hat{n}$ is comparable to or larger than the mean error itself. The resulting error surfaces over the $(N_{\mathrm{obs}}, \sigma)$ plane for each benchmark, together with the observation-type ablation for 1D, constitute the benchmark reported in Section~3.

\section{Results and Discussion}
\label{sec:results}

The inverse PINN framework described in Section~\ref{sec:methods} is evaluated on the two benchmark problems through the structured sparsity--noise sweep, the observation-type ablation, and the multi-seed trajectory analysis. Results for the one-dimensional MacDonald benchmark are presented in Section~\ref{subsec:results_1d}; results for the two-dimensional sloped channel are presented in Section~\ref{subsec:results_2d}; the comparison between the two benchmarks and the broader implications of the findings are discussed in Section~\ref{subsec:discussion}.

\subsection{One-dimensional MacDonald benchmark}
\label{subsec:results_1d}

\subsubsection{Forward verification at the representative configuration}

\begin{figure}[h!]
\centering
\includegraphics[width=0.95\textwidth]{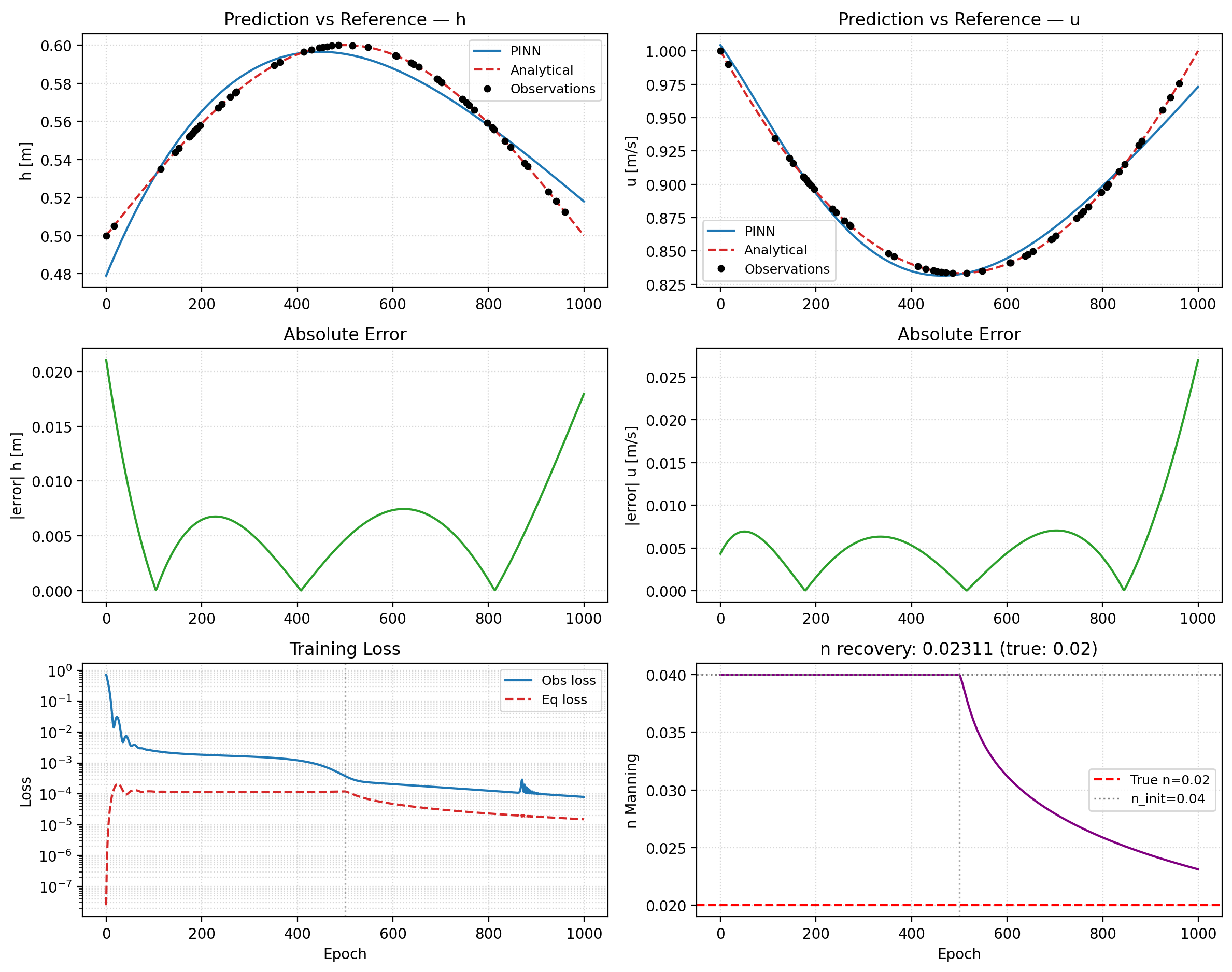}
\caption{Forward verification of the trained 1D inverse PINN at the representative configuration ($N_{\mathrm{obs}} = 50$, no noise, seed $0$). Top row: depth and velocity profiles, with the analytical reference (dashed), the network prediction (solid), and the observation points (markers). Middle row: pointwise absolute error. Bottom row: training-loss history and the friction-coefficient trajectory, showing the Phase~A plateau and the Phase~B descent.}
\label{fig:forward_1d}
\end{figure}

The representative configuration used for verification of the trained network uses $N_{\mathrm{obs}} = 50$ noiseless observations of both depth and velocity, drawn uniformly at random from the reference grid. Figure~\ref{fig:forward_1d} summarizes the outcome of a single training run at this configuration. The trained network reproduces the analytical depth and velocity profiles to within absolute pointwise errors below $0.02$\,m and $0.03$\,m/s respectively, with the largest deviations concentrated near the domain boundaries. The recovered Manning coefficient is $\hat{n} = 0.02311$ against a target $n_{\mathrm{true}} = 0.020$, corresponding to a relative error of $15.5\%$. The loss history exhibits the expected two-phase structure: in Phase~A, only the observation loss decreases while the friction parameter remains at its initial value of $0.04$; at the phase transition near epoch $500$, the physics residual decouples from the constant-$n$ regime and the friction parameter begins to descend monotonically toward its converged value.

\subsubsection{Sparsity and noise sensitivity}

\begin{figure}[h!]
\centering
\includegraphics[width=0.95\textwidth]{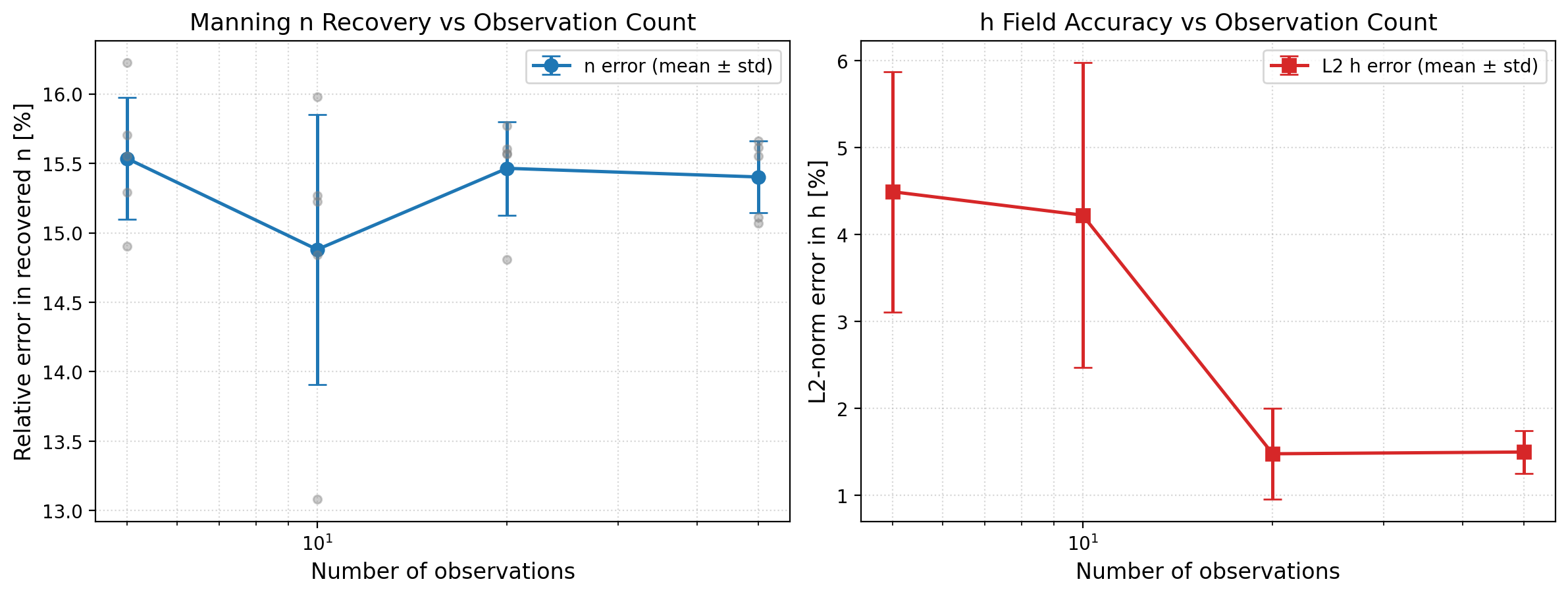}
\caption{1D MacDonald benchmark: dependence of recovered friction coefficient (left) and depth-field reconstruction error (right) on the number of observations at zero noise. Markers and error bars indicate mean and standard deviation over five seeds; grey dots show individual seed values.}
\label{fig:sparsity_1d}
\end{figure}

Figure~\ref{fig:sparsity_1d} shows the dependence of the recovered friction coefficient and the depth-field accuracy on the number of observations at zero noise, with statistics taken over five independent random seeds. The recovered $n$ exhibits a relative error of approximately $15\%$ that is essentially independent of the number of observations across the tested range from five to fifty points. In contrast, the depth-field reconstruction error decreases markedly with additional observations, falling from $4.5\%$ at five points to $1.5\%$ at fifty points and saturating thereafter. This decoupling is informative: additional observations improve the network's representation of the flow field but do not further constrain the friction parameter, indicating that the residual error in $\hat{n}$ at this benchmark is not driven by observation scarcity but by an intrinsic property of the one-dimensional configuration discussed in Section~\ref{subsec:discussion}.

\begin{figure}[h!]
\centering
\includegraphics[width=0.95\textwidth]{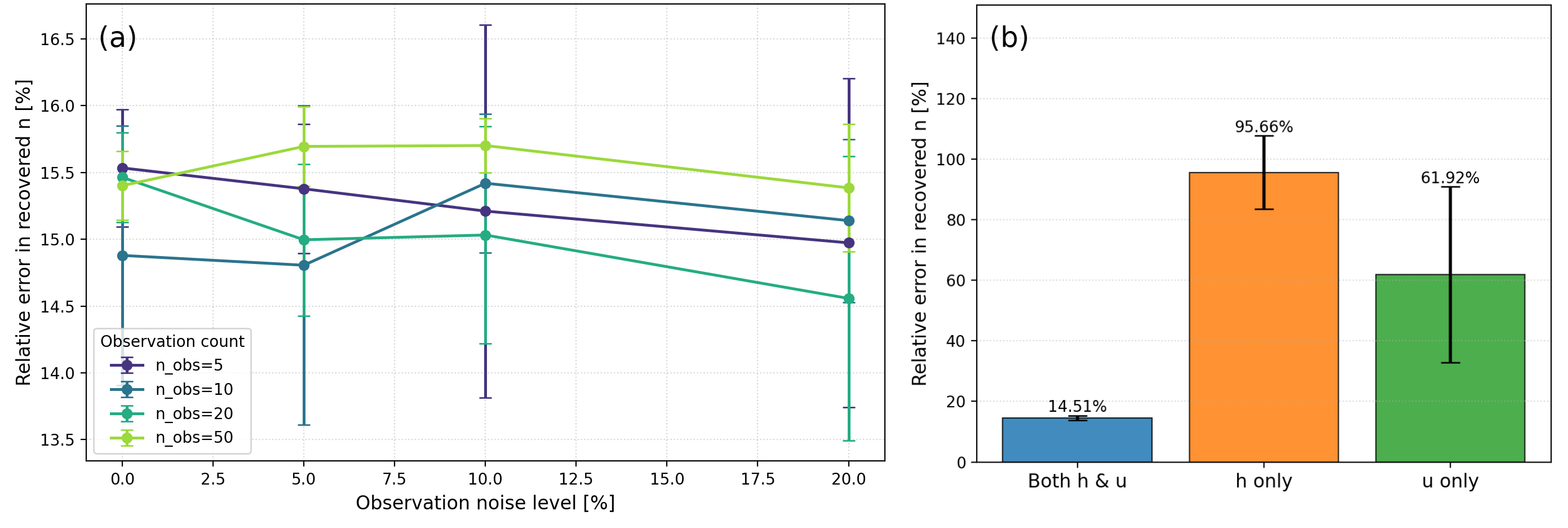}
\caption{1D MacDonald benchmark: (a) relative error in the recovered friction coefficient as a function of observation noise level, for each observation count. Markers and error bars indicate mean and standard deviation over five seeds; (b) observation-type ablation at $N_{\mathrm{obs}} = 20$ and no noise. Bars show seed-mean relative error in the recovered friction coefficient with standard-deviation error bars over five seeds.}
\label{fig:noise_obs_1d}
\end{figure}

The dependence on observation noise, shown in Figure~\ref{fig:noise_obs_1d}a, is consistent with the sparsity result. The relative error in $\hat{n}$ remains in the narrow range $14.6$--$15.7\%$ across all sixteen combinations of observation count and noise level tested in the full sparsity--noise grid, including the largest noise magnitude of $20\%$ of the field standard deviation. The recovered friction coefficient is therefore insensitive to both axes of variation, indicating that the residual recovery error is structural rather than data-driven.

\subsubsection{Observation-type ablation}

To isolate the contribution of each observed variable to friction identifiability, the observation-type ablation was performed at the fixed grid point $N_{\mathrm{obs}} = 20$, no noise, with the observed variable set restricted in turn to depth only, velocity only, and both. The results, shown in Figure~\ref{fig:noise_obs_1d}b, reveal a strong asymmetry. When both depth and velocity are observed, the relative error in $\hat{n}$ is $14.5\%$, consistent with the sparsity and noise sweeps. When only the depth is observed, the recovery fails catastrophically, with a mean relative error of $95.7\%$ and tight seed-to-seed agreement, indicating a systematic collapse of the friction parameter rather than seed-dependent failure. When only the velocity is observed, the mean relative error is $61.9\%$ with substantially larger seed variability, indicating that the recovery becomes practically non-identifiable in this regime. This ablation is a direct empirical demonstration that joint observation of depth and velocity is necessary for friction identifiability in the one-dimensional shallow-water inverse problem: neither variable alone provides sufficient information to constrain $n$, regardless of how densely it is sampled.

\subsubsection{Seed-to-seed reproducibility}

\begin{figure}[h!]
\centering
\includegraphics[width=0.65\textwidth]{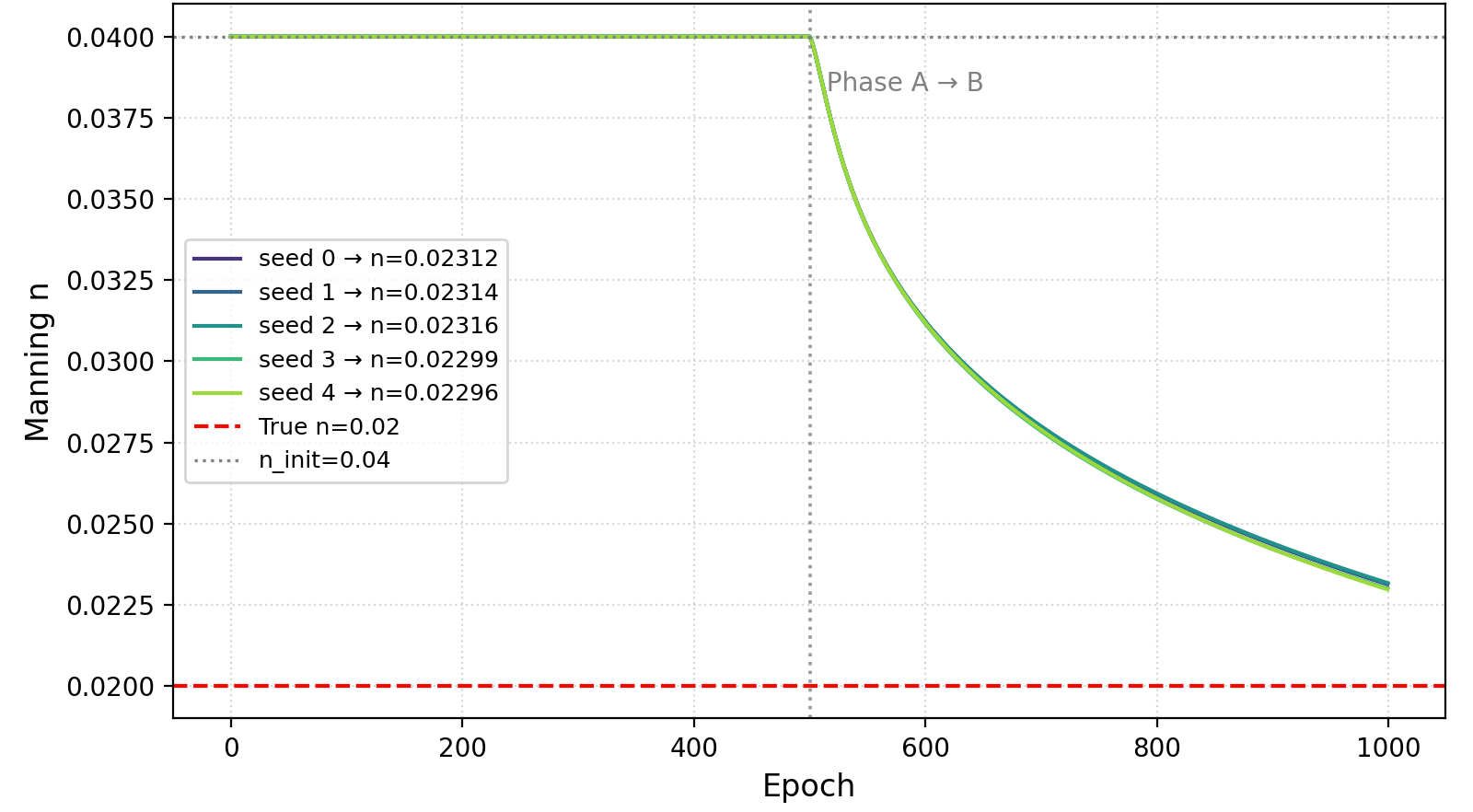}
\caption{1D MacDonald benchmark: convergence trajectories of the Manning coefficient across five seeds at $N_{\mathrm{obs}} = 20$ and no noise. The vertical dotted line marks the transition from Phase~A to Phase~B.}
\label{fig:trajectory_1d}
\end{figure}

Figure~\ref{fig:trajectory_1d} shows the trajectory of the recovered friction coefficient across five seeds for a representative configuration ($N_{\mathrm{obs}} = 20$, no noise). All five trajectories remain at the initial value $n_{\mathrm{init}} = 0.04$ throughout Phase~A and follow nearly identical descent curves in Phase~B, converging to recovered values in the narrow range $0.02296$--$0.02316$. The seed-to-seed variability is two orders of magnitude smaller than the bias from $n_{\mathrm{true}} = 0.020$, confirming that the residual recovery error is dominated by a systematic component rather than stochastic variability across runs.

\subsection{Two-dimensional sloped channel benchmark}
\label{subsec:results_2d}

\subsubsection{Test case geometry}

\begin{figure}[h!]
\centering
\includegraphics[width=0.95\textwidth]{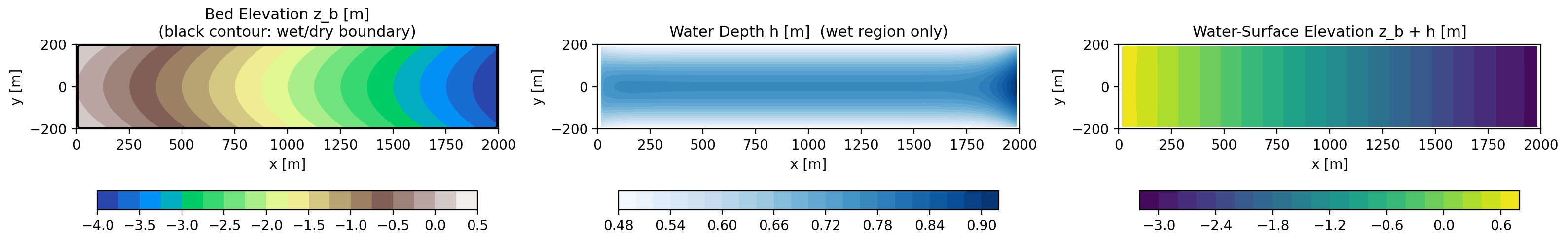}
\caption{2D sloped-channel benchmark: bed elevation (left), water depth (centre), and water-surface elevation (right) at steady state, obtained from the finite-volume solver with $n_{\mathrm{true}} = 0.020$.}
\label{fig:geometry_2d}
\end{figure}

Figure~\ref{fig:geometry_2d} shows the geometry of the two-dimensional benchmark generated by the finite-volume solver described in Section~\ref{subsec:benchmarks}. The bed elevation decreases monotonically in the streamwise direction and increases parabolically toward the lateral walls, producing a centred flow channel of approximately $4{,}600$ wet cells. The water depth ranges from $0.48$\,m near the lateral walls to $0.93$\,m at the channel centreline. The depth and velocity fields are smooth, and the transverse velocity component, though small in magnitude, is non-zero throughout the interior due to the lateral curvature of the bed.

\subsubsection{Forward verification at the representative configuration}

\begin{figure}[h!]
\centering
\includegraphics[width=0.98\textwidth]{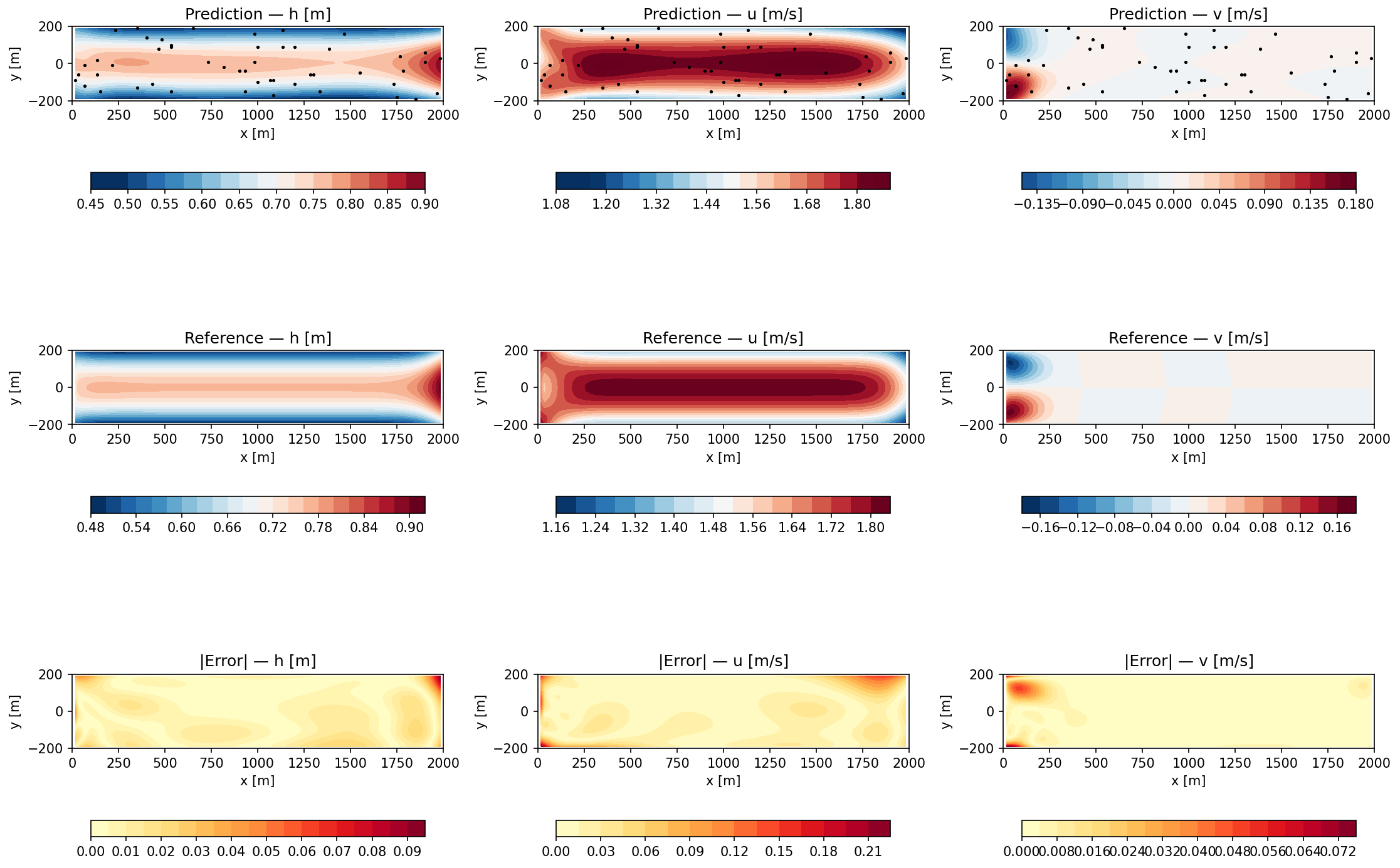}
\caption{Forward verification of the trained 2D inverse PINN at the representative configuration ($N_{\mathrm{obs}} = 50$, no noise, seed $0$). Top row: predicted depth, streamwise velocity, and transverse velocity fields with observation locations overlaid. Middle row: reference fields from the finite-volume solver. Bottom row: absolute pointwise error.}
\label{fig:forward_2d}
\end{figure}

The representative configuration for the two-dimensional benchmark uses $N_{\mathrm{obs}} = 50$ noiseless observations of depth and both velocity components. The forward verification of the trained network at this configuration is shown in Figure~\ref{fig:forward_2d}. The predicted depth and streamwise velocity fields reproduce the reference solution across the interior of the domain, with absolute errors below $0.05$\,m in depth and $0.1$\,m/s in streamwise velocity over most of the wet domain; localized error concentrations appear at the inflow boundary, where Dirichlet conditions in the reference solver introduce a sharp adjustment that the smooth network representation cannot fully resolve. The recovered Manning coefficient is $\hat{n} = 0.02087$ against a target of $0.020$, corresponding to a relative error of $4.4\%$.

\begin{figure}[h!]
\centering
\includegraphics[width=0.95\textwidth]{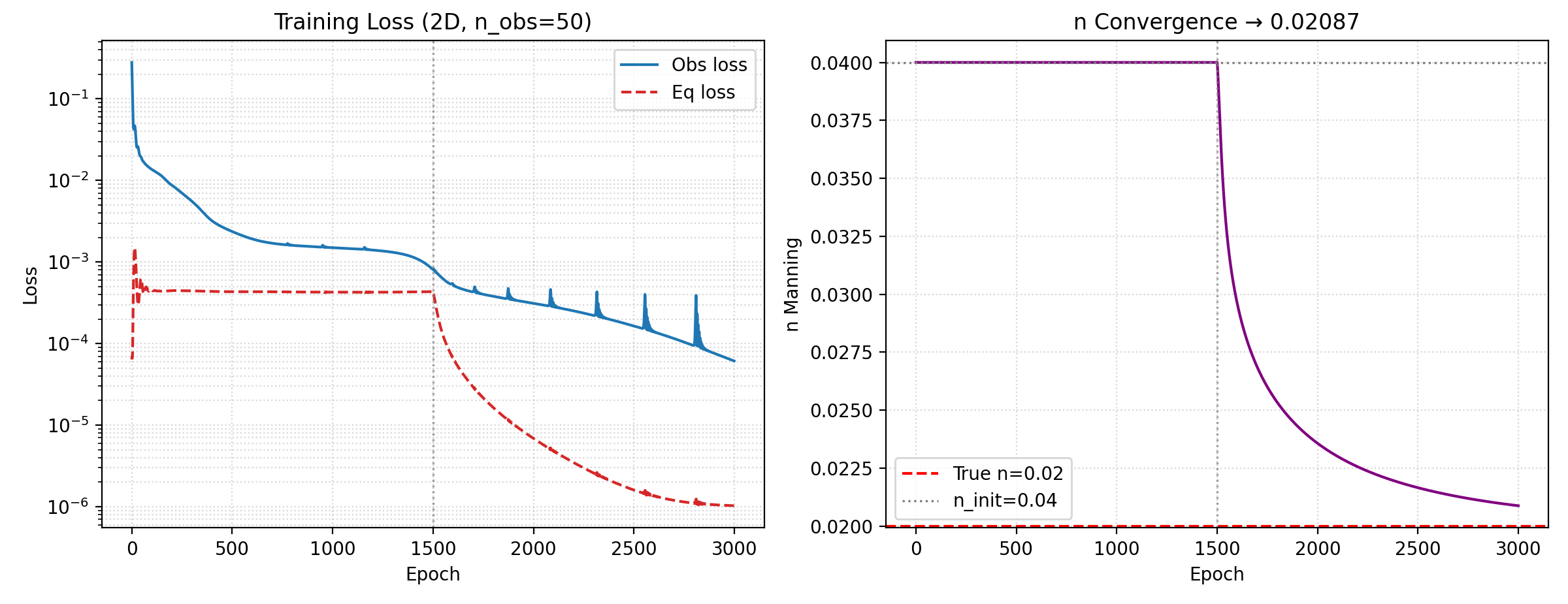}
\caption{2D sloped-channel benchmark: training diagnostics at the representative configuration. Left: observation and physics loss as a function of epoch, with the Phase~A to Phase~B transition marked by the vertical dotted line. Right: convergence trajectory of the Manning coefficient toward its target value.}
\label{fig:convergence_2d}
\end{figure}

The associated training diagnostics in Figure~\ref{fig:convergence_2d} confirm that the two-phase schedule behaves as designed: the observation loss decreases by roughly four orders of magnitude during Phase~A while the friction parameter remains at its initial value, after which Phase~B activates the physics residual and drives the friction parameter from $0.04$ down to its converged value of $0.02087$ over the subsequent $1500$ steps.

\subsubsection{Sparsity and noise sensitivity}

\begin{figure}[h!]
\centering
\includegraphics[width=0.95\textwidth]{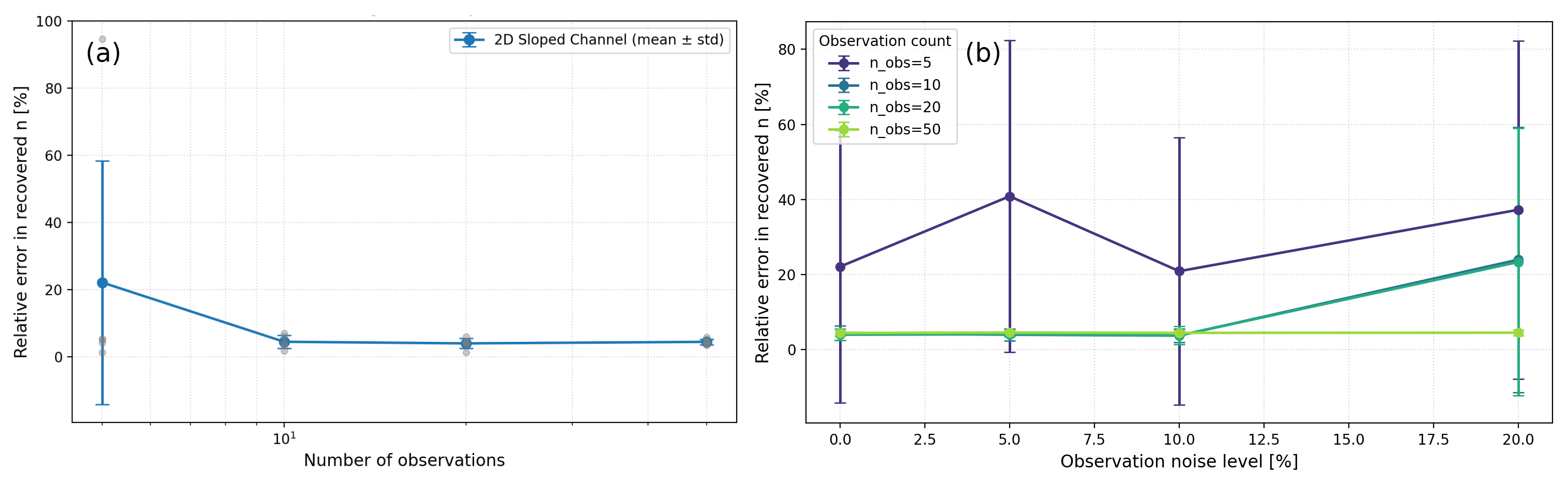}
\caption{2D sloped-channel benchmark: (a) dependence of the recovered friction coefficient on the number of observations at zero noise. Markers and error bars show mean and standard deviation over five seeds; grey dots show individual seed values. (b) relative error in the recovered friction coefficient as a function of observation noise level, for each observation count. Markers and error bars show mean and standard deviation over five seeds.}
\label{fig:sparsity_2d}
\end{figure}

Figure~\ref{fig:sparsity_2d}a shows the dependence of the recovered friction coefficient on the number of observations at zero noise. A sharp transition is observed between $N_{\mathrm{obs}} = 5$ and $N_{\mathrm{obs}} = 10$: at the sparsest configuration, the seed-mean relative error is $22.1\%$ with a seed standard deviation of approximately $36\%$, indicating that recovery is essentially unreliable at this density. From $N_{\mathrm{obs}} = 10$ onward, the recovery stabilizes at a relative error of approximately $4\%$ with seed-to-seed variability below $2\%$. The plateau persists through $N_{\mathrm{obs}} = 50$, indicating that ten observations are sufficient to identify the friction coefficient in this two-dimensional configuration and that additional observations yield diminishing returns.

\begin{figure}[h!]
\centering
\includegraphics[width=0.65\textwidth]{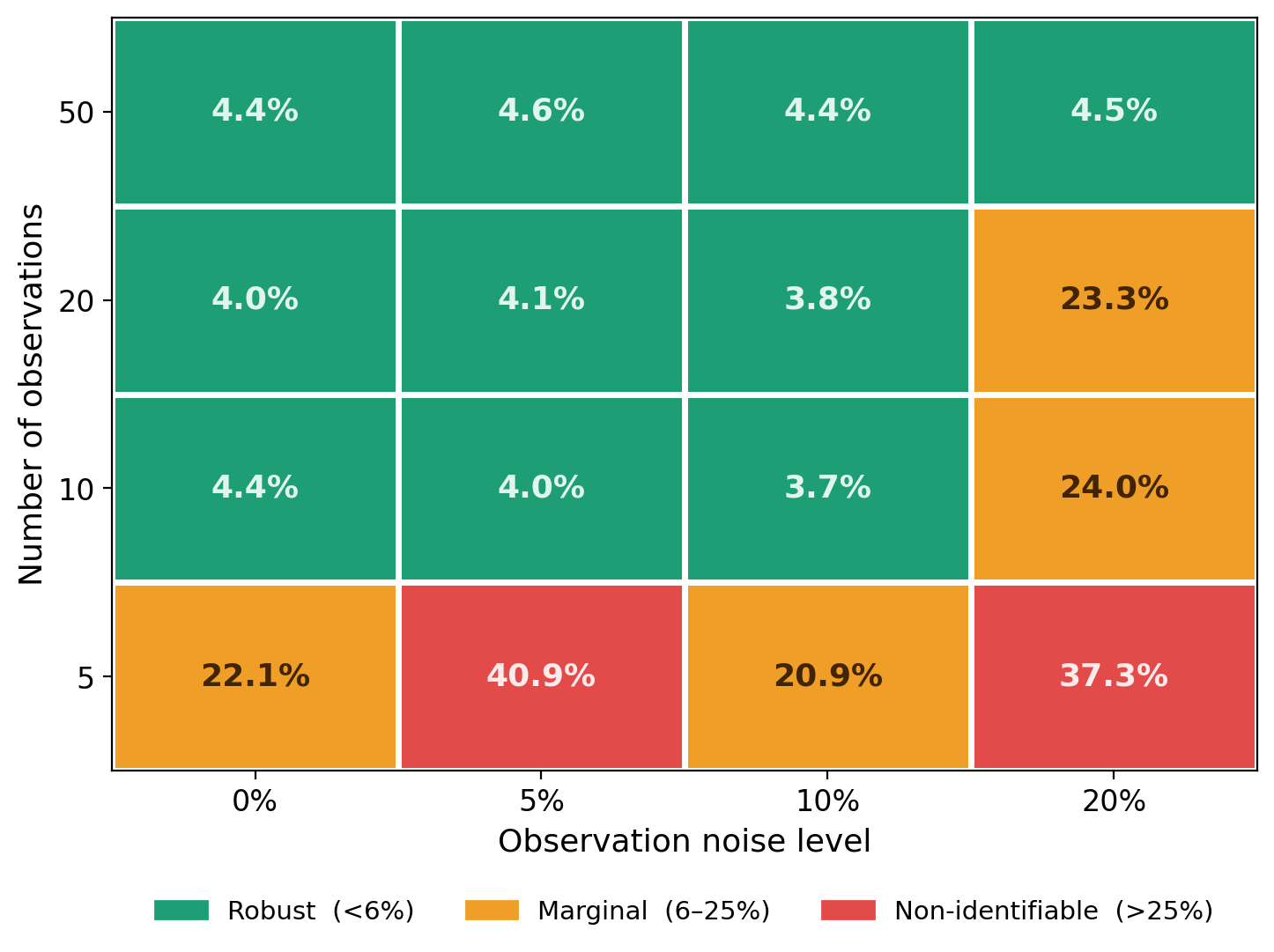}
\caption{2D sloped-channel benchmark: relative error in the recovered friction coefficient across the full sparsity--noise grid. Cell values are seed-mean relative errors in percent.}
\label{fig:heatmap_2d}
\end{figure}

The dependence on observation noise, shown in Figure~\ref{fig:sparsity_2d}b, reveals a sparsity-dependent noise tolerance. For $N_{\mathrm{obs}} = 50$, the recovery is robust to noise up to and including $20\%$ of the field standard deviation, with the error remaining near $4.5\%$ across the noise range. For $N_{\mathrm{obs}} = 10$ and $N_{\mathrm{obs}} = 20$, the recovery is robust up to $10\%$ noise but degrades sharply at $20\%$ noise, where the error grows to approximately $23\%$--$24\%$ with substantially increased seed-to-seed variability. For $N_{\mathrm{obs}} = 5$, the recovery is unreliable at all noise levels. The full sparsity--noise grid is summarized in the heatmap in Figure~\ref{fig:heatmap_2d}, which displays a clear two-region structure: a robust-recovery region with errors below $5\%$ spanning $N_{\mathrm{obs}} \geq 10$ and noise levels up to $10\%$, and a failure region at the lowest observation density or the highest noise level.

\subsubsection{Seed-to-seed reproducibility}

\begin{figure}[h!]
\centering
\includegraphics[width=0.65\textwidth]{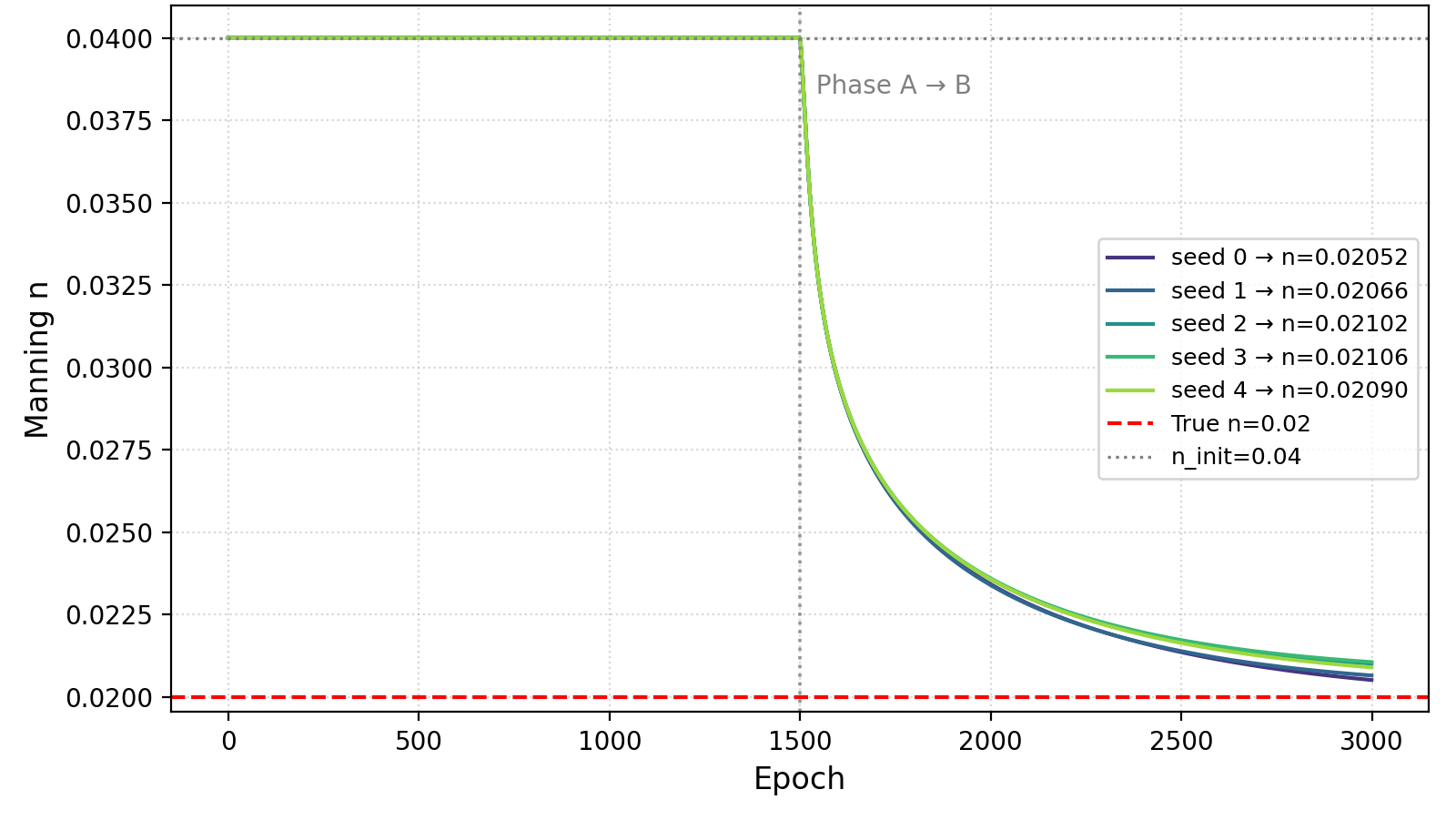}
\caption{2D sloped-channel benchmark: convergence trajectories of the Manning coefficient across five seeds at $N_{\mathrm{obs}} = 20$ and no noise. The vertical dotted line marks the transition from Phase~A to Phase~B.}
\label{fig:trajectory_2d}
\end{figure}

The convergence trajectories of the friction coefficient across five seeds at $N_{\mathrm{obs}} = 20$ and zero noise are shown in Figure~\ref{fig:trajectory_2d}. All five trajectories descend from $n_{\mathrm{init}} = 0.04$ to converged values in the range $0.02052$--$0.02106$, corresponding to relative errors between $2.6\%$ and $5.3\%$. The trajectories are visually indistinguishable for most of the training, and the converged values agree to within $0.0005$ in absolute terms, confirming that the recovery is highly reproducible across random initializations in this configuration.

\subsection{Summary tables and cross-benchmark comparison}
\label{subsec:discussion}

A summary of the recovery results at zero noise is given in Table~\ref{tab:summary}, which reports the seed-mean recovered friction coefficient, its standard deviation, the relative error in $\hat{n}$, and the depth-field $L_{2}$ error for both benchmarks at each tested observation count. The observation-type ablation results for the one-dimensional case are reported separately in Table~\ref{tab:ablation}.

\begin{table}[h!]
\centering
\caption{Recovered Manning coefficient and depth-field accuracy at zero noise for both benchmarks. Statistics are mean $\pm$ standard deviation over five seeds. Target value: $n_{\mathrm{true}} = 0.020$.}
\label{tab:summary}
\begin{tabular}{lcccc}
\hline
Benchmark & $N_{\mathrm{obs}}$ & $\hat{n}$ & $\varepsilon_{n}$ [\%] & $L_{2,h}$ [\%] \\
\hline
1D MacDonald      &  5 & $0.02310 \pm 0.00009$ & $15.52 \pm 0.43$  & $4.49 \pm 1.41$   \\
                  & 10 & $0.02298 \pm 0.00020$ & $14.88 \pm 0.97$  & $4.22 \pm 1.74$   \\
                  & 20 & $0.02309 \pm 0.00007$ & $15.46 \pm 0.34$  & $1.59 \pm 0.61$   \\
                  & 50 & $0.02308 \pm 0.00005$ & $15.40 \pm 0.26$  & $1.59 \pm 0.36$   \\
\hline
2D sloped channel &  5 & $0.02442 \pm 0.00728$ & $22.11 \pm 36.4$  & $19.61 \pm 12.78$ \\
                  & 10 & $0.02087 \pm 0.00033$ & $4.36 \pm 1.65$   & $8.84 \pm 1.55$   \\
                  & 20 & $0.02083 \pm 0.00025$ & $4.02 \pm 1.27$   & $6.54 \pm 2.97$   \\
                  & 50 & $0.02087 \pm 0.00018$ & $4.44 \pm 0.91$   & $3.09 \pm 0.39$   \\
\hline
\end{tabular}
\end{table}

\begin{table}[h!]
\centering
\caption{Observation-type ablation for the 1D MacDonald benchmark at $N_{\mathrm{obs}} = 20$ and no noise. Statistics are mean $\pm$ standard deviation over five seeds.}
\label{tab:ablation}
\begin{tabular}{lcc}
\hline
Observed variables & $\hat{n}$ & $\varepsilon_{n}$ [\%] \\
\hline
Depth and velocity ($h$, $u$) & $0.02290 \pm 0.00008$ & $14.51 \pm 0.41$ \\
Depth only ($h$)              & $0.00087 \pm 0.00248$ & $95.66 \pm 12.4$ \\
Velocity only ($u$)           & $0.00761 \pm 0.05818$ & $61.92 \pm 29.2$ \\
\hline
\end{tabular}
\end{table}

Three observations stand out from the comparison between the two benchmarks summarized in Table~\ref{tab:summary}. First, the two-dimensional benchmark achieves substantially lower error than the one-dimensional benchmark at every observation count from ten upward, with errors near $4\%$ in 2D against approximately $15\%$ in 1D. Second, the two-dimensional benchmark exhibits a clear identifiability threshold near $N_{\mathrm{obs}} = 10$, below which the seed-to-seed variability becomes comparable to the recovered value itself, while the one-dimensional benchmark shows essentially flat performance across the tested range. Third, in both benchmarks, the recovered values are consistently biased above the target $n_{\mathrm{true}} = 0.020$ rather than scattered symmetrically around it.

A fourth point of agreement between the two benchmarks emerges in the depth-field accuracy reported in Table~\ref{tab:summary}: in both cases, the $L_{2,h}$ error continues to decrease with additional observations even after the recovered friction coefficient has plateaued. This decoupling between field reconstruction and parameter recovery reinforces the interpretation that the residual bias in $\hat{n}$ is structural rather than driven by observation scarcity.

The lower error in the two-dimensional benchmark, despite its higher nominal complexity, is interpreted as a consequence of the richer information content of each two-dimensional observation. Each observation in the two-dimensional case provides three measured quantities (depth and two velocity components) whereas each one-dimensional observation provides only two. More importantly, the transverse velocity component arises specifically from the lateral curvature of the bed combined with the friction-controlled momentum balance, and its magnitude scales directly with the friction coefficient through the cross-momentum equation. This makes the transverse velocity an informative signal for friction identifiability that has no analogue in the one-dimensional case, and is consistent with the observation-type ablation in 1D, which already showed that even within a single dimension, removing one velocity component degrades the recovery catastrophically.

The systematic positive bias observed in both benchmarks, in which the recovered friction coefficient overestimates the target by a fixed fractional amount that is insensitive to observation density or noise, is interpreted as an identifiability bias intrinsic to the joint-optimization formulation. Within Phase~B, the network is free to deform the predicted flow field within the slack permitted by the finite observation density, and the bias-variance trade-off between the network's flexibility and the scalar friction parameter resolves in favour of a slightly elevated friction coefficient that balances residuals across the domain. The fact that the bias is approximately three times larger in the one-dimensional benchmark than in the two-dimensional benchmark is consistent with this interpretation: the additional constraint provided by the transverse momentum equation in 2D narrows the slack available to the network and pulls the recovered friction closer to its true value.

The observation-type ablation in 1D provides a definitive identifiability statement that is independent of the bias magnitude: joint observation of depth and velocity is necessary for any meaningful recovery, and either variable alone is insufficient. This result has direct implications for the design of observation networks in practical applications, where it argues that co-located depth and velocity measurements at a small number of points are substantially more informative than dense measurements of either variable alone.

\section{Conclusions}
\label{sec:conclusions}

This study presented a systematic benchmarking framework for the recovery of the Manning friction coefficient from sparse, noisy observations of the shallow water equations using inverse physics-informed neural networks. Two canonical configurations were considered: the one-dimensional MacDonald subcritical channel, in which the bed elevation was back-solved from the steady momentum balance to produce a self-consistent analytical reference, and a two-dimensional sloped channel with a parabolic transverse cross-section, for which the steady reference field was generated by a well-balanced finite-volume solver. The recovery accuracy was mapped over a structured grid of observation sparsity, observation noise, and observation type, with multi-seed statistics used to distinguish robust regimes from configurations in which the inverse problem becomes practically non-identifiable. The main findings are as follows:

\begin{enumerate}[label=(\alph*)]
\item The inverse PINN is capable of recovering the Manning friction coefficient from sparse and noisy observations in both one- and two-dimensional configurations. In the two-dimensional benchmark, recovery errors remain below 5\% when at least 10 point observations of depth and both velocity components are available.

\item The recovery exhibits a clear identifiability threshold at low observation densities. In the two-dimensional case, the transition from unreliable recovery with five observations to stable recovery with ten observations is abrupt and is clearly visible in the sparsity--noise heatmap.

\item Noise tolerance depends strongly on the number of observations. With 50 observations, the inverse PINN remains robust up to 20\% noise relative to the field standard deviation, whereas 10 to 20 observations remain robust up to approximately 10\% noise before recovery degrades.

\item Joint observation of depth and velocity is essential for friction identifiability. The one-dimensional observation-type ablation shows that recovery from depth-only or velocity-only observations fails severely and cannot be corrected simply by increasing the number of measurements.

\item The two-dimensional benchmark consistently outperforms the one-dimensional benchmark in absolute recovery accuracy. This improvement is attributed to the additional information carried by the transverse velocity component, whose magnitude is directly linked to the friction coefficient through the cross-momentum equation.

\item The application of a second-order quasi-Newton refinement after Adam was found to be actively harmful for the inverse problem, as it exploits a near-null direction in the joint parameter--network landscape and degrades the recovered friction coefficient even while reducing both loss components further.

\item The two-phase training schedule, in which the physics residual is activated only after the network has been pre-conditioned on observations, was found to be essential for avoiding a collapse of the friction parameter toward zero in the early optimization.

\item The choice of how loss weights are stored in the compiled training graph was found to have a decisive effect on whether the friction parameter receives any gradient updates at all, despite leaving no visible signature in the loss curves.

\end{enumerate}

\section*{Code and Data Availability}

All codes and data required to reproduce the benchmark experiments, figures, and tables are available at \url{https://github.com/sradfar/inversePINN}.

\bibliographystyle{apalike}
\bibliography{references}

@article{Raissi2019,
  author  = {Raissi, Maziar and Perdikaris, Paris and Karniadakis, George Em},
  title   = {Physics-informed neural networks: A deep learning framework for solving forward and inverse problems involving nonlinear partial differential equations},
  journal = {Journal of Computational Physics},
  volume  = {378},
  pages   = {686--707},
  year    = {2019},
  doi     = {10.1016/j.jcp.2018.10.045}
}

@article{Karniadakis2021,
  author  = {Karniadakis, George Em and Kevrekidis, Ioannis G. and Lu, Lu and Perdikaris, Paris and Wang, Sifan and Yang, Liu},
  title   = {Physics-informed machine learning},
  journal = {Nature Reviews Physics},
  volume  = {3},
  pages   = {422--440},
  year    = {2021},
  doi     = {10.1038/s42254-021-00314-5}
}

@article{Lu2021,
  author  = {Lu, Lu and Meng, Xuhui and Mao, Zhiping and Karniadakis, George Em},
  title   = {{DeepXDE}: A deep learning library for solving differential equations},
  journal = {SIAM Review},
  volume  = {63},
  number  = {1},
  pages   = {208--228},
  year    = {2021},
  doi     = {10.1137/19M1274067}
}

@article{Cai2021,
  author  = {Cai, Shengze and Wang, Zhicheng and Wang, Sifan and Perdikaris, Paris and Karniadakis, George Em},
  title   = {Physics-informed neural networks for heat transfer problems},
  journal = {Journal of Heat Transfer},
  volume  = {143},
  number  = {6},
  pages   = {060801},
  year    = {2021},
  doi     = {10.1115/1.4050542}
}

@article{Cuomo2022,
  author  = {Cuomo, Salvatore and Schiano Di Cola, Vincenzo and Giampaolo, Fabio and Rozza, Gianluigi and Raissi, Maziar and Piccialli, Francesco},
  title   = {Scientific machine learning through physics-informed neural networks: Where we are and what's next},
  journal = {Journal of Scientific Computing},
  volume  = {92},
  pages   = {88},
  year    = {2022},
  doi     = {10.1007/s10915-022-01939-z}
}

@article{Baydin2018,
  author  = {Baydin, Atilim Gunes and Pearlmutter, Barak A. and Radul, Alexey Andreyevich and Siskind, Jeffrey Mark},
  title   = {Automatic differentiation in machine learning: A survey},
  journal = {Journal of Machine Learning Research},
  volume  = {18},
  number  = {153},
  pages   = {1--43},
  year    = {2018}
}

@article{Jagtap2020ConservativePINN,
  author  = {Jagtap, Ameya D. and Kharazmi, Ehsan and Karniadakis, George Em},
  title   = {Conservative physics-informed neural networks on discrete domains for conservation laws: Applications to forward and inverse problems},
  journal = {Computer Methods in Applied Mechanics and Engineering},
  volume  = {365},
  pages   = {113028},
  year    = {2020},
  doi     = {10.1016/j.cma.2020.113028}
}

@article{Jagtap2020AdaptiveActivation,
  author  = {Jagtap, Ameya D. and Kawaguchi, Kenji and Karniadakis, George Em},
  title   = {Adaptive activation functions accelerate convergence in deep and physics-informed neural networks},
  journal = {Journal of Computational Physics},
  volume  = {404},
  pages   = {109136},
  year    = {2020},
  doi     = {10.1016/j.jcp.2019.109136}
}

@article{Jagtap2022OceanEngineering,
  author  = {Jagtap, Ameya D. and Mitsotakis, Dimitrios and Karniadakis, George Em},
  title   = {Deep learning of inverse water waves problems using multi-fidelity data: Application to Serre--Green--Naghdi equations},
  journal = {Ocean Engineering},
  volume  = {248},
  pages   = {110775},
  year    = {2022},
  doi     = {10.1016/j.oceaneng.2022.110775}
}

@article{Jin2021NSFnet,
  author  = {Jin, Xiaowei and Cai, Shengze and Li, Hui and Karniadakis, George Em},
  title   = {{NSFnets}: Navier--Stokes flow nets for physics-informed fluid flow reconstruction and equations discovery},
  journal = {Journal of Computational Physics},
  volume  = {426},
  pages   = {109951},
  year    = {2021},
  doi     = {10.1016/j.jcp.2020.109951}
}

@article{Wang2021GradientPathologies,
  author  = {Wang, Sifan and Teng, Yujun and Perdikaris, Paris},
  title   = {Understanding and mitigating gradient flow pathologies in physics-informed neural networks},
  journal = {SIAM Journal on Scientific Computing},
  volume  = {43},
  number  = {5},
  pages   = {A3055--A3081},
  year    = {2021},
  doi     = {10.1137/20M1318043}
}

@article{Wang2021EigenvectorBias,
  author  = {Wang, Sifan and Wang, Hanwen and Perdikaris, Paris},
  title   = {On the eigenvector bias of Fourier feature networks: From regression to solving multi-scale PDEs with physics-informed neural networks},
  journal = {Computer Methods in Applied Mechanics and Engineering},
  volume  = {384},
  pages   = {113938},
  year    = {2021},
  doi     = {10.1016/j.cma.2021.113938}
}

@inproceedings{Krishnapriyan2021,
  author    = {Krishnapriyan, Aditi S. and Gholami, Amir and Zhe, Shandian and Kirby, Robert M. and Mahoney, Michael W.},
  title     = {Characterizing possible failure modes in physics-informed neural networks},
  booktitle = {Advances in Neural Information Processing Systems},
  volume    = {34},
  pages     = {26548--26560},
  year      = {2021}
}

@book{Toro2001,
  author    = {Toro, Eleuterio F.},
  title     = {Shock-Capturing Methods for Free-Surface Shallow Flows},
  publisher = {John Wiley \& Sons},
  address   = {Chichester, UK},
  year      = {2001}
}

@book{LeVeque2002,
  author    = {LeVeque, Randall J.},
  title     = {Finite Volume Methods for Hyperbolic Problems},
  publisher = {Cambridge University Press},
  address   = {Cambridge, UK},
  year      = {2002},
  doi       = {10.1017/CBO9780511791253}
}

@book{Chaudhry2008,
  author    = {Chaudhry, M. Hanif},
  title     = {Open-Channel Flow},
  edition   = {2},
  publisher = {Springer},
  address   = {New York},
  year      = {2008},
  doi       = {10.1007/978-0-387-68648-6}
}

@article{Bates2010,
  author  = {Bates, Paul D. and Horritt, Matthew S. and Fewtrell, Timothy J.},
  title   = {A simple inertial formulation of the shallow water equations for efficient two-dimensional flood inundation modelling},
  journal = {Journal of Hydrology},
  volume  = {387},
  number  = {1--2},
  pages   = {33--45},
  year    = {2010},
  doi     = {10.1016/j.jhydrol.2010.03.027}
}

@article{Neal2012,
  author  = {Neal, Jeffrey and Schumann, Guy and Bates, Paul},
  title   = {A subgrid channel model for simulating river hydraulics and floodplain inundation over large and data sparse areas},
  journal = {Water Resources Research},
  volume  = {48},
  number  = {11},
  pages   = {W11506},
  year    = {2012},
  doi     = {10.1029/2012WR012514}
}

@book{Tarantola2005,
  author    = {Tarantola, Albert},
  title     = {Inverse Problem Theory and Methods for Model Parameter Estimation},
  publisher = {SIAM},
  address   = {Philadelphia, PA},
  year      = {2005},
  doi       = {10.1137/1.9780898717921}
}

@book{Chow1959,
  author    = {Chow, Ven Te},
  title     = {Open-Channel Hydraulics},
  publisher = {McGraw-Hill},
  address   = {New York},
  year      = {1959}
}

@techreport{Arcement1989,
  author      = {Arcement, George J. and Schneider, Verne R.},
  title       = {Guide for Selecting Manning's Roughness Coefficients for Natural Channels and Flood Plains},
  institution = {U.S. Geological Survey},
  type        = {Water-Supply Paper},
  number      = {2339},
  year        = {1989},
  address     = {Washington, DC}
}

@book{Cunge1980,
  author    = {Cunge, Jean A. and Holly, Forrest M. and Verwey, Adri},
  title     = {Practical Aspects of Computational River Hydraulics},
  publisher = {Pitman},
  address   = {London},
  year      = {1980}
}

@article{Horritt2002,
  author  = {Horritt, Matthew S. and Bates, Paul D.},
  title   = {Evaluation of 1D and 2D numerical models for predicting river flood inundation},
  journal = {Journal of Hydrology},
  volume  = {268},
  number  = {1--4},
  pages   = {87--99},
  year    = {2002},
  doi     = {10.1016/S0022-1694(02)00121-X}
}

@article{Sampson2015,
  author  = {Sampson, Christopher C. and Smith, Andrew M. and Bates, Paul D. and Neal, Jeffrey C. and Alfieri, Lorenzo and Freer, Jim E.},
  title   = {A high-resolution global flood hazard model},
  journal = {Water Resources Research},
  volume  = {51},
  number  = {9},
  pages   = {7358--7381},
  year    = {2015},
  doi     = {10.1002/2015WR016954}
}

@article{Beven2006,
  author  = {Beven, Keith},
  title   = {A manifesto for the equifinality thesis},
  journal = {Journal of Hydrology},
  volume  = {320},
  number  = {1--2},
  pages   = {18--36},
  year    = {2006},
  doi     = {10.1016/j.jhydrol.2005.07.007}
}

@article{MacDonald1997,
  author  = {MacDonald, Ian and Baines, Michael J. and Nichols, Nick K. and Samuels, Peter G.},
  title   = {Analytic benchmark solutions for open-channel flows},
  journal = {Journal of Hydraulic Engineering},
  volume  = {123},
  number  = {11},
  pages   = {1041--1045},
  year    = {1997},
  doi     = {10.1061/(ASCE)0733-9429(1997)123:11(1041)}
}

@article{Delestre2013,
  author  = {Delestre, Olivier and Lucas, Carine and Ksinant, Pierre-Antoine and Darboux, Frédéric and Laguerre, Christian and Vo, Tinh-Tinh and James, François and Cordier, Stéphane},
  title   = {{SWASHES}: A compilation of shallow water analytic solutions for hydraulic and environmental studies},
  journal = {International Journal for Numerical Methods in Fluids},
  volume  = {72},
  number  = {3},
  pages   = {269--300},
  year    = {2013},
  doi     = {10.1002/fld.3741}
}

@article{Chen2023CoastalEngineering,
  author  = {Chen, Qin and Wang, Nan and Chen, Zhao},
  title   = {Simultaneous mapping of nearshore bathymetry and waves based on physics-informed deep learning},
  journal = {Coastal Engineering},
  volume  = {183},
  pages   = {104337},
  year    = {2023},
  doi     = {10.1016/j.coastaleng.2023.104337}
}

@article{Huang2025EAAI,
  author  = {Huang, Xinyu and Tang, Jun and Shen, Yongming and Zhao, Yanlong and Hao, Shuai},
  title   = {A physically informed neural network approach for modeling wave transformation in vegetated waters},
  journal = {Engineering Applications of Artificial Intelligence},
  volume  = {159},
  pages   = {111803},
  year    = {2025},
  doi     = {10.1016/j.engappai.2025.111803}
}

@article{Ehlers2024Fluids,
  author  = {Ehlers, Svenja and Wagner, Niklas A. and Scherzl, Annamaria and Klein, Marco and Hoffmann, Norbert and Stender, Merten},
  title   = {Data Assimilation and Parameter Identification for Water Waves Using the Nonlinear Schrödinger Equation and Physics-Informed Neural Networks},
  journal = {Fluids},
  volume  = {9},
  number  = {10},
  pages   = {231},
  year    = {2024},
  doi     = {10.3390/fluids9100231}
}

@article{Ehlers2025PRF,
  author  = {Ehlers, Svenja and Hoffmann, Norbert and Tang, Tianning and Callaghan, Adrian H. and Cao, Rui and Padilla, Enrique M. and Fang, Yuxin and Stender, Merten},
  title   = {Physics-informed neural networks for phase-resolved data assimilation and prediction of nonlinear ocean waves},
  journal = {Physical Review Fluids},
  volume  = {10},
  number  = {9},
  pages   = {094901},
  year    = {2025},
  doi     = {10.1103/ytyy-pvys}
}

@article{Mohammadi2025NonlinearDynamics,
  author  = {Mohammadi, Nima and Abbaszadeh, Mostafa and Dehghan, Mehdi and Heitzinger, Clemens},
  title   = {Parameter identification of shallow water waves using the generalized equal width equation and physics-informed neural networks: A conservative approximation scheme},
  journal = {Nonlinear Dynamics},
  volume  = {113},
  pages   = {6491--6516},
  year    = {2025},
  doi     = {10.1007/s11071-024-10497-y}
}

@article{Hu2025JOE,
  author  = {Hu, Shuang and Liu, Meiqin and Zhang, Senlin and Dong, Shanling and Zheng, Ronghao},
  title   = {Physics-Informed Neural Networks for Modeling Ocean Dynamics and Parameter Estimation: Leveraging Ocean Reanalysis Data},
  journal = {IEEE Journal of Oceanic Engineering},
  volume  = {50},
  number  = {3},
  pages   = {2248--2260},
  year    = {2025},
  doi     = {10.1109/JOE.2025.3538927}
}

@article{Frei2026LOM,
  author  = {Frei, S. and Munir, M. U. and Seiverth, A. M. and Gilfedder, B. S.},
  title   = {Using physics-informed neural networks to quantify submarine groundwater discharge under high-frequency tidal dynamics using heat as a tracer},
  journal = {Limnology and Oceanography: Methods},
  volume  = {24},
  pages   = {e70015},
  year    = {2026},
  doi     = {10.1002/lom3.70015}
}

@article{Chen2026JMSE,
  author  = {Chen, Weiyun and Tao, Hairong and Wang, Lei and Fan, Shaofen},
  title   = {Frequency-Domain Physics-Informed Neural Networks for Modeling and Parameter Inversion of Wave-Induced Seabed Response},
  journal = {Journal of Marine Science and Engineering},
  volume  = {14},
  number  = {8},
  pages   = {690},
  year    = {2026},
  doi     = {10.3390/jmse14080690}
}

@article{Someya2025GJI,
  author  = {Someya, Masayoshi and Furumura, Takashi},
  title   = {Physics-informed neural networks for offshore tsunami data assimilation},
  journal = {Geophysical Journal International},
  volume  = {242},
  number  = {3},
  pages   = {ggaf243},
  year    = {2025},
  doi     = {10.1093/gji/ggaf243}
}

@misc{SolverLoop2026,
  author        = {Cardoso-Bihlo, Elsa and Bihlo, Alex},
  title         = {A solver-in-the-loop framework for end-to-end differentiable coastal hydrodynamics},
  year          = {2026},
  eprint        = {2604.07129},
  archivePrefix = {arXiv},
  primaryClass  = {physics.flu-dyn}
}

@article{Holman2013,
  author  = {Holman, Robert A. and Plant, Nathaniel G. and Holland, K. Todd},
  title   = {{cBathy}: A robust algorithm for estimating nearshore bathymetry},
  journal = {Journal of Geophysical Research: Oceans},
  volume  = {118},
  number  = {5},
  pages   = {2595--2609},
  year    = {2013},
  doi     = {10.1002/jgrc.20199}
}

@article{Wilson2014,
  author  = {Wilson, Gregory W. and Ozkan-Haller, H. Tuba and Holman, Robert A. and Haller, Merrick C. and Honegger, David A. and Chickadel, C. Chris},
  title   = {Surf zone bathymetry and circulation predictions via data assimilation of remote sensing observations},
  journal = {Journal of Geophysical Research: Oceans},
  volume  = {119},
  number  = {3},
  pages   = {1993--2016},
  year    = {2014},
  doi     = {10.1002/2013JC009213}
}

@article{Salim2021,
  author  = {Salim, Salam and Wilson, Gregory W.},
  title   = {Data assimilation of nearshore bathymetry using wave height and alongshore current observations},
  journal = {Coastal Engineering},
  volume  = {170},
  pages   = {104009},
  year    = {2021},
  doi     = {10.1016/j.coastaleng.2021.104009}
}

@article{Medeiros2012,
  author  = {Medeiros, Stephen C. and Hagen, Scott C. and Weishampel, John F.},
  title   = {Comparison of floodplain surface roughness parameters derived from land cover data and field measurements},
  journal = {Journal of Hydrology},
  volume  = {452--453},
  pages   = {139--149},
  year    = {2012},
  doi     = {10.1016/j.jhydrol.2012.05.043}
}

@article{Dyakonova2018,
  author  = {Dyakonova, Tatyana and Khoperskov, Alexander},
  title   = {Bottom friction models for shallow water equations: Manning's roughness coefficient and small-scale bottom heterogeneity},
  journal = {Journal of Physics: Conference Series},
  volume  = {973},
  pages   = {012032},
  year    = {2018},
  doi     = {10.1088/1742-6596/973/1/012032}
}

@article{Radfar2023WACE,
  author  = {Radfar, Soheil and Galiatsatou, Panagiota and Wahl, Thomas},
  title   = {Application of nonstationary extreme value analysis in the coastal environment -- A systematic literature review},
  journal = {Weather and Climate Extremes},
  volume  = {41},
  pages   = {100575},
  year    = {2023},
  doi     = {10.1016/j.wace.2023.100575}
}

@article{Radfar2023OceanEngineering,
  author  = {Radfar, Soheil and Galiatsatou, Panagiota},
  title   = {Influence of nonstationarity and dependence of extreme wave parameters on the reliability assessment of coastal structures -- A case study},
  journal = {Ocean Engineering},
  volume  = {273},
  pages   = {113862},
  year    = {2023},
  doi     = {10.1016/j.oceaneng.2023.113862}
}

@article{Maghsoodifar2025IJDRR,
  author  = {Maghsoodifar, Faezeh and Radfar, Soheil and Moftakhari, Hamed and Moradkhani, Hamid},
  title   = {Establishing closure criteria for coastal roadways under flooding conditions},
  journal = {International Journal of Disaster Risk Reduction},
  volume  = {130},
  pages   = {105805},
  year    = {2025},
  doi     = {10.1016/j.ijdrr.2025.105805}
}

@misc{Radfar2025ALPINE,
  author        = {Radfar, Soheil and Maghsoodifar, Faezeh and Moftakhari, Hamed and Moradkhani, Hamid},
  title         = {Integrating Newton's Laws with Deep Learning for Enhanced Physics-Informed Compound Flood Modeling},
  year          = {2025},
  eprint        = {2507.15021},
  archivePrefix = {arXiv},
  primaryClass  = {physics.geo-ph}
}

@article{Radfar2026HESS,
  author  = {Radfar, Soheil and Moftakhari, Hamed and Mu{\~n}oz, David F. and Gori, Avantika and Diermanse, Ferdinand and Lin, Ning and AghaKouchak, Amir},
  title   = {Towards a typology for hybrid compound flood modeling},
  journal = {Hydrology and Earth System Sciences},
  volume  = {30},
  pages   = {1397--1420},
  year    = {2026},
  doi     = {10.5194/hess-30-1397-2026}
}

@article{Radfar2026ERL,
  author  = {Radfar, Soheil and Taheri, Parastoo and Moftakhari, Hamed},
  title   = {Global evidence for co-variability of tidal constituents},
  journal = {Environmental Research Letters},
  volume  = {21},
  pages   = {024019},
  year    = {2026},
  doi     = {10.1088/1748-9326/ae2c0c}
}

@article{Mahmoudi2025EarthsFuture,
  author  = {Mahmoudi, Sadaf and Moftakhari, Hamed and Mu{\~n}oz, David F. and Radfar, Soheil and Sweet, William and Moradkhani, Hamid},
  title   = {Escalating High Tide Flooding Along the Atlantic and Gulf Coast of the United States Due To Sea Level Rise},
  journal = {Earth's Future},
  volume  = {13},
  pages   = {e2024EF005328},
  year    = {2025},
  doi     = {10.1029/2024EF005328}
}

@article{Wahl2015,
  author  = {Wahl, Thomas and Jain, Shaleen and Bender, Jens and Meyers, Steven D. and Luther, Mark E.},
  title   = {Increasing risk of compound flooding from storm surge and rainfall for major US cities},
  journal = {Nature Climate Change},
  volume  = {5},
  pages   = {1093--1097},
  year    = {2015},
  doi     = {10.1038/nclimate2736}
}

@article{Moftakhari2017,
  author  = {Moftakhari, Hamed R. and AghaKouchak, Amir and Sanders, Brett F. and Matthew, Richard A.},
  title   = {Compounding effects of sea level rise and fluvial flooding},
  journal = {Proceedings of the National Academy of Sciences},
  volume  = {114},
  number  = {37},
  pages   = {9785--9790},
  year    = {2017},
  doi     = {10.1073/pnas.1620325114}
}

@article{Zscheischler2018,
  author  = {Zscheischler, Jakob and Westra, Seth and van den Hurk, Bart J. J. M. and Seneviratne, Sonia I. and Ward, Philip J. and Pitman, Andy and AghaKouchak, Amir and Bresch, David N. and Leonard, Michael and Wahl, Thomas and Zhang, Xuebin},
  title   = {Future climate risk from compound events},
  journal = {Nature Climate Change},
  volume  = {8},
  pages   = {469--477},
  year    = {2018},
  doi     = {10.1038/s41558-018-0156-3}
}

@article{audusse2004fast,
  title={A fast and stable well-balanced scheme with hydrostatic reconstruction for shallow water flows},
  author={Audusse, Emmanuel and Bouchut, Fran{\c{c}}ois and Bristeau, Marie-Odile and Klein, Rupert and Perthame, Beno{\i}⁁ t},
  journal={SIAM Journal on Scientific Computing},
  volume={25},
  number={6},
  pages={2050--2065},
  year={2004},
  publisher={SIAM}
}

@article{radfar2024nature,
  title={Nature-based solutions as buffers against coastal compound flooding: Exploring potential framework for process-based modeling of hazard mitigation},
  author={Radfar, Soheil and Mahmoudi, Sadaf and Moftakhari, Hamed and Meckley, Trevor and Bilskie, Matthew V and Collini, Renee and Alizad, Karim and Cherry, Julia A and Moradkhani, Hamid},
  journal={Science of The Total Environment},
  volume={938},
  pages={173529},
  year={2024},
  publisher={Elsevier}
}

@article{wang2026unraveling,
  title={Unraveling uncertainty in compound flood modeling: sensitivity of simulations to forcings and model parameters},
  author={Wang, Chen and Gomez, Francisco J and Mosavat, Mohammad and Radfar, Soheil and Moftakhari, Hamed and Moradkhani, Hamid},
  journal={Journal of Hydrology},
  pages={135424},
  year={2026},
  publisher={Elsevier}
}

\end{document}